\pdfoutput=1
\RequirePackage{fix-cm}
\documentclass[3p,twocolumn,number,sort&compress]{elsarticle}
\usepackage{microtype}
\usepackage{graphicx}
\usepackage{csquotes}
\usepackage{caption}
\usepackage{subcaption}
\usepackage{xcolor}
\usepackage{blindtext}
\usepackage{siunitx}
\usepackage{amsmath}
\usepackage{tabulary}
\DeclareMathOperator{\erfc}{erfc}
\DeclareSIUnit\atoms{atoms}
\DeclareSIUnit\ions{ions}
\DeclareSIUnit\monolayer{\SI{E15}{atoms}}
\DeclareSIUnit\years{a}
\DeclareSIUnit\counts{cts}
\DeclareSIUnit\countspersecond{cps}
\DeclareSIUnit\hydroxygroup{-OH-groups}
\usepackage{a4wide}
\usepackage{fancyhdr}
\usepackage{needspace}

\makeatletter
\def\ps@pprintTitle{%
	\let\@oddhead\@empty
	\let\@evenhead\@empty
	\def\@oddfoot{}%
	\let\@evenfoot\@oddfoot}
\makeatother

\begin{document}

\begin{frontmatter}
\title{Alpha spectrometric characterization of thin \textsuperscript{233}U sources for \textsuperscript{229(m)}Th production}

\author[1,2,3,4]{Raphael Haas}
\author[1]{Michelle Hufnagel}
\author[1]{Roman Abrosimov}
\author[1,2,3,4]{Christoph E. D\"ullmann}
\author[5]{Dominik Krupp}
\author[1,2]{Christoph Mokry}
\author[1,2]{Dennis Renisch}
\author[1,3]{Jörg Runke}
\author[5]{Ulrich W. Scherer}

\address[1]{Department of Chemistry - TRIGA Site, Johannes Gutenberg University Mainz, Mainz, Germany}
\address[2]{Helmholtz Institute Mainz, Mainz, Germany}
\address[3]{GSI Helmholtzzentrum für Schwerionenforschung GmbH, Darmstadt, Germany}
\address[4]{PRISMA Cluster of Excellence, Johannes Gutenberg University Mainz, Mainz, Germany}
\address[5]{Institut für Physikalische Chemie und Radiochemie, Hochschule Mannheim - University of Applied Sciences, 68163 Mannheim, Germany}

\begin{abstract}
	Four different techniques were applied for the production of \textsuperscript{233}U alpha recoil ion sources, providing \textsuperscript{229}Th ions. They were compared with respect to a minimum energy spread of the \textsuperscript{229}Th recoil ions, using the emitted alpha particles as an indicator. The techniques of Molecular Plating, Drop-on-Demand inkjet printing, chelation from dilute nitric acid solution on chemically functionalized silicon surfaces, and self-adsorption on passivated titanium surfaces were used. All fabricated sources were characterized by using alpha spectrometry, radiographic imaging, and scanning electron microscopy. A direct validation for the estimated recoil ion rate was obtained by collecting \textsuperscript{228}Th recoil ions from \textsuperscript{232}U recoil ion sources prepared by self-adsorption and Molecular Plating. The chelation and the self-adsorption based approaches appear most promising for the preparation of recoil ion sources delivering monochromatic recoil ions.

\end{abstract}
\end{frontmatter}

\section{Introduction}\label{Intro}

Radioactive sources of alpha-decaying or spontaneously fissioning radionuclides are used in many experiments in nuclear chemistry and nuclear physics as a source for their daughter nuclei. In alpha-decaying recoil ion sources, the daughter nuclei are emitted from thin active layers owing to the short residual range of the daughters in matter. Such recoil ion sources were recently used in physics experiments, e.g., for investigations of the nuclear clock isomer \textsuperscript{229m}Th \cite{Wense2016,Thielking2018,Seiferle2019} and for optimizations of the extraction of short-lived isotopes in buffer gas cells \cite{Droese2014,Goetz2020}. The \textsuperscript{233}U is used as example nuclide in the source fabrication, as it is of high current interest due to the production of \textsuperscript{229m}Th and will be used as a recoil ion source in the experiment of the TACTICa\footnote{Trapping And Cooling of Thorium Ions with Calcium} collaboration \cite{Grootberning2019,Stopp2019,Haas2020}. In the TACTICa experiment, the \textsuperscript{229(m)}Th recoil ions will be decelerated by an electric field to less than \SI{1}{\kilo\electronvolt} kinetic energy to allow their trapping in a linear Paul trap. In contrast to the approach employed, e.g. in \cite{Wense2016,Thielking2018,Seiferle2019}, where recoil ions were thermalized in buffer gas, stringent vacuum requirements and the desire to retain the initial charge state distribution \cite{Haas2020} render purely electrostatic deceleration favorable for TACTICa. Ideal recoil ion sources for application in TACTICa consist of a monolayer of the mother nuclide \textsuperscript{233}U to gain a high recoil efficiency with a minimal energy dispersion of the recoil ions. Due to the approximately fixed areal density of a monolayer, the activity of such an ion source is defined by the active area.

Different methods exist to fabricate targets or radioactive sources in many different geometries and thicknesses. A well established method is the Molecular Plating (MP) technique \cite{Parker1962}. This method was well characterized and enhanced by Vascon et al. to fabricate smooth crack-free large-area thin films of alpha particle emitting \textsuperscript{147}Sm \cite{Vascon2013}. These layers were investigated with X-ray photoelectron spectroscopy (XPS) and alpha spectrometry and showed that specific MP parameters, like the choice of plating solvent and the roughness of the substrate, lead to effects which influence the recoil efficiency and energy dispersion of the alpha-recoil ions. The thickness and morphology of the produced salt layer as well as cracked solvent fragments deposited on top of the radioactive source proved to influence the relative alpha detection efficiency by as much as \SI{15}{\percent} \cite{Vascon2013a}. Such alpha detection losses imply a critical influence on the recoil efficiency of the alpha daughter ions and their energy distribution. The MP technique needs further developments to fabricate more efficient recoil ion sources.

A novel fabrication technique is the Drop-on-Demand ink-jet printing method (DoD) \cite{Haas2017}, which is able to deposit dissolved material onto a variety of solid surfaces and is not restricted to electrically conducting substrates. It has a high printing accuracy as well as a reliable reproducibility of the deposited material \cite{Haas2017}. This method has not yet been used to fabricate recoil ion sources consisting of a few atomic layers. Printing as well as solution parameters are well adjustable and, therefore, it should be possible to produce recoil ion sources suitable for application in, e.g. the TACTICa experiment.

In contrast to MP and DoD, where the material adheres mainly to the substrate by physisorption, there are two other possible ways to achieve effectively a monolayer of radionuclides, which is chemisorbed to the substrate. Krupp et al. developed a method to functionalize silicon surfaces of, e.g. passivated implanted planar silicon (PIPS) detectors for alpha spectrometry \cite{Krupp2018}. The silicon surfaces are functionalized with sulfonic acid groups, which chelate radionuclides from an acid solution. The chelation creates a chemisorption of the radionuclides, so that the surface can be rinsed afterwards with distilled water to remove non-bound radionuclides without loosing bound material. Theoretically, a saturated functionalized surface contains a single layer of radionuclides and seems promising for the production of ideal recoil ion sources.

Another way to produce chemisorbed monolayer films is to make use of the chemical equilibrium of radionuclides in solution on metal oxide surfaces, well known as self-adsorption (SA) process. SA processes are well investigated for many radionuclides on a wide variety of metal oxides and soil material in the field of nuclear waste disposal \cite{Jaffrezic-Renault1980,Wazne2006,Mueller2012,Lamb2016}. Self-adsorption of thorium was used for the production of thin \textsuperscript{229}Th films on CaF\textsubscript{2} surfaces \cite{Yamaguchi2015} and is also promising for the production of ideal recoil ion sources for a high recoil efficiency with small kinetic energy dispersion of the emitted recoil ions.

In this article, the theoretical aspects of recoil ion sources concerning the stopping powers of alpha particles and recoil daughters in the source material as well as the achievable areal density of uranium depending on the fabrication method is discussed. Furthermore, Advanced Alpha spectrometric Simulations (AASI) \cite{Siiskonen2005,Pollanen2006,Siiskonen2011} were performed for the investigation of the influence of different source parameters on the alpha spectra. The experimental and simulated spectra were parameterized using an adapted fit routine \cite{Bortels1987,Pomme2008} for a quantitative evaluation, allowing a comparison of recoil sources fabricated by different methods. At last, the recoil efficiencies were investigated of two sources produced by MP and SA by using the shorter-lived \textsuperscript{232}U (half-life: \SI{68.9}{\years}).

\section{Theoretical aspects}\label{Theory}

\subsection{Stopping power in materials}

The stopping powers of the alpha particles and the recoil daughter ions differ by about one order of magnitude. Furthermore, they depend on the chemical nature of the mother nuclide material. For the example nuclide \textsuperscript{233}U (half-life: \SI{1.59E5}{\years}, E$_\alpha$: \SI{4824}{\kilo\electronvolt} (\SI{82.7}{\percent}), \SI{4783}{\kilo\electronvolt} (\SI{14.9}{\percent}), \SI{4729}{\kilo\electronvolt} (\SI{1.85}{\percent})), the kinetic energy \textit{E}\textsubscript{k} of the \textsuperscript{229}Th ions is about \SI{84}{\kilo\electronvolt} while the alpha particle energy is about \SI{4.8}{\mega\electronvolt} [derived from \textit{E}\textsubscript{k}(\textsuperscript{229}Th) = \textit{Q}\textsubscript{$\alpha$}(\textsuperscript{233}U) - \textit{E}\textsubscript{$\alpha$}(\textsuperscript{233}U), where \textit{Q}\textsubscript{$\alpha$} is the Q-value of the alpha decay and \textit{E}\textsubscript{$\alpha$} the kinetic energy of the alpha particle]. The stopping powers and ranges in different materials can easily be calculated with SRIM-2013 \cite{Ziegler2013}. Due to the fact that the exact uranium compound is only known for DoD-produced sources, but not for the others, a number of different uranium compounds were used for the SRIM simulations. The stopping powers and ranges of both the \textsuperscript{229}Th and the alpha particles in these compounds are given in Table~\ref{tab1} and Table~\ref{tab2}. SRIM defines a monolayer of any material as \SI{E15}{\atoms\per\centi\meter\squared}. Therefore, the stopping powers describe the energetic loss per atomic layer. The data show how much more the \textsuperscript{229}Th recoil ions are influenced by the mother nuclide material with ranges below \SI{500}{\angstrom} compared to the alpha particles with ranges above \SI{10}{\micro\meter} in the same materials. Thus, alpha spectra of \textsuperscript{233}U recoil ion sources cannot give detailed information on the kinetic energy distribution of the recoil ions. Even very thin sources with ten atomic layers of the mother nuclide material decelerate the emitted recoil ions by several \SI{}{\kilo\electronvolt}. For a first semi-quantitative screening, a high-resolution alpha spectrum of an uranium recoil ion source allowing for a detailed evaluation of the peak shape may give a hint about the thickness of the source, which can help determining the recoil efficiency and energy distribution of the thorium ions.\\ 

\begin{table*}[!htb]
	\caption{Stopping powers and ranges of \textsuperscript{229}Th recoil ions in different \textsuperscript{233}U compounds. The initial kinetic energy of the recoil ions is set to \SI{84}{\kilo\electronvolt}. The data were obtained with SRIM-2013 \cite{Ziegler2013}.}
	\label{tab1}
	\centering
	\begin{tabulary}{1.0\textwidth}{ C | @{}C@{} | @{}C@{} | C }
		\hline
		Compound & density / \SI{}{\gram\per\cubic\centi\meter} \cite{Grenthe2006,Lide2003crc} & Stopping Power of \textsuperscript{229}Th / \SI{}{\electronvolt\per(\monolayer\per\centi\meter\squared)} & Range / \SI{}{\nano\meter} \\
		\hline
		\parbox[c]{3.5cm}{UO$_2$(OH)$_2$$\times$H$_2$O} & 5.0 & 348 & 28 \\
		UO$_2$CO$_3$ & 5.7 & 510 & 25 \\
		UO$_2$(NO$_3$)$_2$ & 2.8 & 457 & 45 \\
		UO$_2$ & 11.0 & 731 & 15 \\
		U$_3$O$_8$ & 8.4 & 664 & 19 \\
		\hline
	\end{tabulary}
\end{table*}

\begin{table*}[!htb]
	\caption{Stopping powers and ranges of alpha particles in different \textsuperscript{233}U compounds. The maximum kinetic energy of the alpha particles is set to \SI{4.824}{\mega\electronvolt}. The data were obtained with SRIM-2013 \cite{Ziegler2013}.}
	\label{tab2}
	\centering
	\begin{tabulary}{1.0\textwidth}{ C | @{}C@{} | @{}C@{} | C }
		\hline
		Compound & density / \SI{}{\gram\per\cubic\centi\meter} \cite{Grenthe2006,Lide2003crc} & Stopping Power of alpha particle / \SI{}{\electronvolt\per(\monolayer\per\centi\meter\squared)} & Range / \SI{}{\micro\meter} \\
		\hline
		\parbox[c]{3.5cm}{UO$_2$(OH)$_2$$\times$H$_2$O} & 5.0 & 20.6 & 17.3 \\
		UO$_2$CO$_3$ & 5.7 & 29.5 & 15.6 \\
		UO$_2$(NO$_3$)$_2$ & 2.8 & 27.5 & 28.9 \\
		UO$_2$ & 11.0 & 43.1 & 10.7 \\
		U$_3$O$_8$ & 8.4 & 38.9 & 13.1 \\
		\hline
	\end{tabulary}
\end{table*}

\subsection{Theoretical areal density}

As previously mentioned, a single layer of any material is defined in SRIM-2013 as \SI{E15}{\atoms\per\centi\meter\squared}, which is too crude for an approximation of all elements in one layer in order to be directly applicable in the present case. As an example, an areal density of about \SI{2.23E14}{\atoms\per\centi\meter\squared} is calculated for a monolayer of uranium atoms in uranium dioxide. The areal uranium densities of this and other materials are given in Table~\ref{tab3}. As the true chemical species and crystal structure is difficult to investigate for sources prepared by MP and DoD, a value of \SI{E14}{\atoms\per\centi\meter\squared} will be used for further discussions and for the calculation of the concentration of solutions as used in the experimental section. The estimation of an areal uranium density gets more complicated for methods like chelation by sulfonic acid groups and SA, because there the areal uranium densities depend on the density and availability of binding partners on the surface of the substrate.

\begin{table}[!htb]
	\caption{Calculated areal uranium densities in various compounds. Basis of the calculations are crystal structure data from \cite{Grenthe2006,Lide2003crc}.}
	\label{tab3}
	\centering
	\begin{tabulary}{1.0\linewidth}{ C@{\hspace*{1cm}} | @{}C@{} }
		\hline
		\parbox[c]{3.5cm}{Compound} & areal uranium density / \SI{E14}{\atoms\per\centi\meter\squared} \\
		\hline
		UO$_2$(OH)$_2$$\times$H$_2$O & 1.41  \\
		UO$_2$CO$_3$ & 0.32 \\
		UO$_2$(NO$_3$)$_2$ & 0.51  \\
		UO$_2$ & 2.23  \\
		U$_3$O$_8$ & 0.93  \\
		\hline
	\end{tabulary}
\end{table}

For the chelation method, the areal density of sulfonic acid groups is important. An areal uranium density of about \SI{4E15}{\atoms\per\centi\meter\squared} is achievable on PIPS detectors functionalized in that way \cite{Krupp2018}. For SA, the density of oxygen binding partners on the surface of the substrate influences the amount of adsorbed uranium atoms. Investigations in the field of nuclear waste disposal have shown the highest adsorption rate of uranium for titanium dioxide \cite{Lamb2016}. These investigations were always performed with titanium dioxide colloids, whereas source fabrication is performed on a titanium foil, which provides a lower density of oxygen binding partners than colloids do. To increase the source activity, the density of these binding partners can be increased, e.g., by passivation. There are several methods to achieve titanium passivation, e.g., by anodic oxidation or by heating \cite{Velten2001}. Thermal treatment allows investigating the adsorption yield (and thus the achievable areal uranium density) by specifically producing one of two obtainable titanium dioxide modifications. This is carried out by the irreversible transformation of anatase$\rightarrow$rutile at temperatures >\SI{600}{\degreeCelsius} at atmospheric pressure \cite{Xiliang2009,Velten2001}. The exact transformation temperature depends on impurities in the titanium. D. Velten et al. found transformations for amorphous$\rightarrow$anatase and anatase$\rightarrow$rutile in sol-gel powder samples after heat treatment \cite{Velten2001}. Anatase, the modification found naturally on a passivation layer of titanium, is the most promising species for high adsorption yields. The areal oxygen density of both, anatase and rutile, was calculated for some ideal crystal faces to derive an average number. Calculations based on crystal structure data \cite{Lide2003crc,Osti_1207597,Osti_1267830} are given in Table~\ref{tab4}. The mean areal density of oxygen atoms on the surface of anatase and rutile is similar and is about \SI{1.3E15}{\atoms\per\centi\meter\squared}. Due to the large ionic radius and coordination sphere of uranyl ions in carbonate solutions \cite{Jaffrezic-Renault1980,Lamb2016}, the uranyl ion will be coordinated to two oxygen atoms on the surface. This leads to an U:O ratio of 1:3 to 1:2, corresponding to a maximum areal uranium density of \SI{6.5E14}{\atoms\per\centi\meter\squared} with SA. The real areal density of adsorbed uranium is probably lower due to defects in the crystal surface. \SI{E15}{\atoms\per\centi\meter\squared} of \textsuperscript{233}U have an activity of \SI{138}{\becquerel\per\centi\meter\squared}, corresponding to an emission rate of \textsuperscript{229}Th ions into a $2\pi$ solid angle of about \SI{69}{\ions\per\second\per\centi\meter\squared}.

\begin{table*}[!htb]
	\caption{Calculated areal densities (ad) of oxygen available for self-adsorption on ideal crystal faces of anatase and rutile based on crystal structure data from \cite{Lide2003crc,Osti_1207597,Osti_1267830}. The Miller indices of the individual crystal faces are given. The areal densities of oxygen are given for each crystal face and as average values for each modification.}
	\label{tab4}
	\centering
	\begin{tabulary}{1.05\textwidth}{ C | C | C | @{}C@{} | @{}C@{} }
		\hline
		& lengths / \SI{}{\angstrom} & crystal face & ad / \SI{E15}{\atoms\per\centi\meter\squared} & mean ad / \SI{E15}{\atoms\per\centi\meter\squared} \\
		\hline
		& a: 3.785 & (001) & 1.40 & \\
		Anatase & c: 9.514 & (100) & 1.11 & 1.18\\
		&  & (101) & 1.02 & \\
		\hline
		& a: 4.594 & (001) & 1.51 & \\
		Rutile & c: 2.962 & (110) & 0.98 & 1.36\\
		&  & (100) & 1.60 & \\
		\hline
	\end{tabulary}
\end{table*}

\section{Peak fit model and AASIFIT}\label{SimFit}

\begin{figure}
	\centering
	\includegraphics[width=1.0\linewidth]{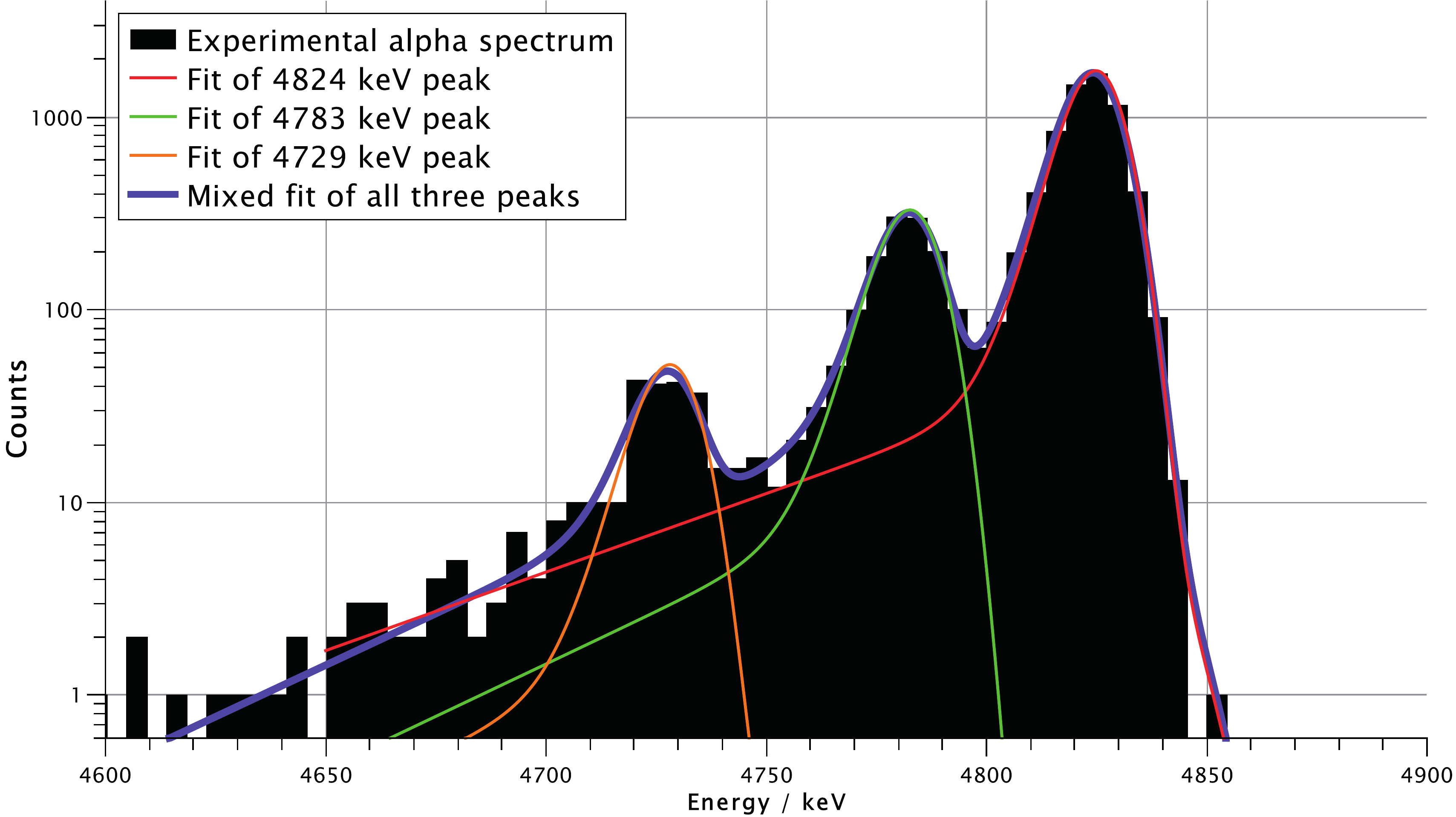}
	\caption{Exemplary QtiPlot fit of a 2k channel alpha spectrum of purified \textsuperscript{233}U. A mixed fit including all three peaks of the multiplett (Eq.~\ref{equation3}) as well as independent fits (Eq.~\ref{equation1}) of the single peaks are shown.}
	\label{fig:1}
\end{figure}

Alpha spectrometry is an important radioanalytical technique for the quantitative and qualitative investigation of alpha-emitting radionuclide samples in environmental and nuclear research \cite{Aggarwal2016}. High-resolution spectra of high-quality sources help characterizing samples with regard to alpha-decaying impurities, isotopic compositions and even qualitative characteristics like thickness and homogeneity of the sample \cite{Pomme2008}. Often, nearby peaks in alpha spectra overlap at least partially. For the deconvolution and interpretation of such spectra, several peak fit models were developed and improved in the last decades. One of the most successful models to represent a mono-energetic alpha peak consists of a Gaussian joined with an exponential distribution \cite{Bortels1987,LHOIR1984}

\begin{equation*}\label{equation1}
\begin{gathered}
f\left(u-\mu ;\sigma ,\tau \right)=\frac{A}{2\tau }\mathrm{exp}\left(\frac{u-\mu}{\tau }+\frac{{\sigma }^{2}}{2{\tau }^{2}}\right)\\
\times\erfc\left[\frac{1}{\sqrt{2}}\left(\frac{u-\mu }{\sigma }+\frac{\sigma }{\tau }\right)\right]
\end{gathered}
\end{equation*}

where $A$ is the peak area, $u$ is the distance to the peak position $\mu$, $\sigma$ is the standard deviation of the Gaussian and $\tau$ is the tailing parameter. This model was improved by subtraction of the long tail distribution and a mix of two exponential functions with different lengths, $\tau_1<\tau_2$, and a normalized weighting factor $\eta$ \cite{Babeliowsky1993}. In the present case of monolayer alpha recoil sources, the peaks are expected to be very narrow and the long tail distribution is expected to be non-existent, in which case it can be neglected. Therefore, equation~\ref{equation1} was modified with the normalized weighting factor $\eta$ and by a sum of three functions for the present nuclide of interest, \textsuperscript{233}U, showing a multiplet of the three significant peaks

\begin{equation*}\label{equation2}
\begin{gathered}
F\left(u\right)=\sum_{i=1}^{2}\sum_{j=1}^{3}{\eta_i}{f_j}\left(u-{\mu_i} ;{\sigma_i} ,{\tau_i}\right) \\
1={\eta_1}+{\eta_2}
\end{gathered}
\end{equation*}

with individual peak positions $\mu_1$, $\mu_2$ and $\mu_3$ and two shared standard deviations of the Gaussian $\sigma$ and tailing parameters $\tau$, each pertaining to all three peaks. This mixed function contains eight free parameters

\begin{equation*}\label{equation3}
\begin{gathered}
F\left(u\right)=\\
\left[{\eta_1}{f_1}\left(u-{\mu_1} ;{\sigma_1} ,{\tau_1}\right)+{(1-\eta_1)}{f_1}\left(u-{\mu_1} ; {\sigma_2} ,{\tau_2}\right)\right]\\
+\left[{\eta_1}{f_2}\left(u-{\mu_2} ;{\sigma_1} ,{\tau_1}\right)+{(1-\eta_1)}{f_2}\left(u-{\mu_2} ; {\sigma_2} ,{\tau_2}\right)\right]\\
+\left[{\eta_1}{f_3}\left(u-{\mu_3} ;{\sigma_1} ,{\tau_1}\right)+{(1-\eta_1)}{f_3}\left(u-{\mu_3} ; {\sigma_2} ,{\tau_2}\right)\right]
\end{gathered}
\end{equation*}

which was used for the determination of the individual, characteristic tailing parameters $\eta_1$ and $\tau_1$ of spectra from \textsuperscript{233}U sources produced by the four different methods. The \textsuperscript{233}U spectra were fitted with QtiPlot \cite{Vasilef2013}. An example of the applied fit function is given in Fig.~\ref{fig:1}.

Furthermore, for the investigation of effects of different source characteristics on the alpha spectrum, e.g., the thickness and RMS roughness, the simulation and fitting software AASIFIT was provided by STUK, Finland \cite{Siiskonen2005,Pollanen2006,Siiskonen2011}. Numerous parameters of the detector setup, the source material and dimensions can be set in the software to give iterative Monte Carlo-simulated alpha spectra. Experimental data can be loaded into the software and can then be fitted on the basis of simulated spectra of specific isotopes \cite{Siiskonen2011}. Effects of source layer thickness, root mean square (RMS) roughness, homogeneity and source material on simulated alpha spectra will be discussed in Section \ref{Discu}.

\section{Experimental}\label{Exp}

\subsection{General instrumentation and methods}\label{InstrMeth}

A light microscope (Bresser LCD Micro, 40x - 1600x magnification) was used for visual inspection of the sources. Scanning electron microscopy (SEM) was performed with a Philips XL 30 for the investigation of the topography of the sources in high resolution. Autoradiographic images (RI) weren taken with a Fujifilm FLA-7000 for information about the spacial distribution of radioactivity. Atomic force microscopy (AFM) was performed with a Semilab DS95 SPM System for the investigation of the topology (root mean square roughness) of the sources. For alpha spectrometry, different PIPS detectors and source-detector-distances were used for quantitative yield determination and for qualitative measurements. These are specified for each method in the following sections. A 16K channel analog-digital-converter (Canberra Nuclear Data ND581 ADC, used in 1K and 2K mode) with amplifier (Canberra 2015A AMP/TSCA) was used for data acquisition. A certified \textsuperscript{241}Am source (Amersham Buchler, \SI{290}{\becquerel} homogeneously distributed on an anodized aluminum disc with \SI{16}{\milli\meter} in diameter and \SI{150}{\micro\meter} thickness) was used for efficiency calibration of the used PIPS detectors.

\subsection{Molecular Plating}\label{MP}

\subsubsection{Reagents and materials}

A stock solution of column separated (in July 2017) \textsuperscript{233}U as uranium nitrate in \SI{0.1}{\mole\per\liter} HNO$_3$ (Merck, \SI{65}{\percent}) with a concentration of \SI{2.02}{\milli\gram\per\milli\liter} was used for the preparation of all solutions. For MP, three solutions with concentrations of \SI{38.9}{\micro\gram\per\milli\liter}, \SI{21.5}{\micro\gram\per\milli\liter}, \SI{5.25}{\micro\gram\per\milli\liter} were prepared by dilution of aliquots from the stock solution in \SI{0.1}{\mole\per\liter} HNO$_3$. Either N,N'-dimethylformamide (DMF) (Merck, \SI{99.8}{\percent}) or a mixture of \SI{10}{\percent} isopropanol (Merck, \SI{99.8}{\percent}) and \SI{90}{\percent} isobutanol (Merck, \SI{99}{\percent}) was used as electrolyte in the plating cells. A palladium foil or a 1-mm thick palladium wire (>\SI{99.98}{\percent}) was used as anode. Silicon wafers (\SI{34}{\milli\meter} in diameter, \SI{300}{\micro\meter} thickness) with a titanium coating (\SI{100}{\nano\meter} thickness) produced by magnetron sputtering as well as titanium foils (\SI{99.6}{\percent}, \SI{25}{\milli\meter} in diameter, \SI{25}{\micro\meter} thickness) were used as substrates. For cleaning of the titanium foils, distilled water, \SI{6}{\mole\per\liter} HCl (Merck, \SI{37}{\percent}), isopropanol and acetone (Merck, \SI{99.5}{\percent}) were used.

\subsubsection{Instrumentation and methods}

For depositions on the titanium-coated silicon wafers, a horizontal cell design of A. Vascon et al. \cite{Vascon2013, Vascon2012} was used in combination with a HV power supply (Heinzinger LNC 3000-10 pos., \SIrange{0}{3000}{\volt}, \SIrange{0}{10}{\milli\ampere}). For depositions on the titanium foils, a vertical cell design \cite{Trautmann1989} was used in combination with a different HV power supply (FUG MCP 350-1250, \SIrange{0}{1250}{\volt}, \SIrange{0}{250}{\milli\ampere}). Alpha spectra were recorded with a \SI{2000}{\milli\meter\squared} PIPS detector (Canberra PD2000-40-300AM) in \SI{5}{\milli\meter} distance for yield determination and a \SI{25}{\milli\meter\squared} PIPS detector (Ortec ULTRA\textsuperscript{TM}, \SI{11}{\kilo\electronvolt} FWHM intrinsic resolution) in \SI{100}{\milli\meter} distance was used for recording high resolution spectra.

\subsubsection{Source fabrication and characterization}\label{MPcharacterization}

The titanium foils were cleaned with \SI{6}{\mole\per\liter} HCl, distilled water, isopropanol and acetone. The coated silicon wafers were cleaned with isopropanol to avoid damages to the titanium coating. An aliquot of \SI{100}{\micro\liter} of the \SI{5.25}{\micro\gram\per\milli\liter} \textsuperscript{233}U solution was mixed with \SI{35}{\milli\liter} DMF using a vortex mixer for deposition on one silicon wafer. Another aliquot of \SI{100}{\micro\liter} of the \SI{34.4}{\micro\gram\per\milli\liter} solution was mixed with \SI{3.5}{\milli\liter} isopropanol and \SI{31.5}{\milli\liter} isobutanol using a vortex mixer for deposition on another silicon wafer. For deposition on the titanium foils, aliquots of \SI{20}{\micro\liter} of the \SI{21.5}{\micro\gram\per\milli\liter} solution were mixed with \SI{1}{\milli\liter} isopropanol and \SI{9}{\milli\liter} isobutanol. MP was performed at constant current of about \SI{0.75}{\milli\ampere\per\centi\meter\squared} for \SIrange{1}{2}{\hour}. After deposition, the sources were carefully dried under an IR lamp in a fume hood. Alpha spectra for yield determination were taken with peak areas of at least \SI{E4}{\counts} to reduce statistical counting errors below \SI{1}{\percent}. Qualitative alpha spectra were taken for several days to obtain adequate statistic. SEM pictures and RI were taken for visual inspection of the sources. Irradiation time for RI was about \SI{2}{\hour}.

\subsection{Drop-on-Demand inkjet printing}\label{DoD}

\subsubsection{Reagents and materials}

An aliquot of the stock solution was diluted to give three solutions with concentrations of \SI{138}{\micro\gram\per\milli\liter}, \SI{74.2}{\micro\gram\per\milli\liter} and \SI{18.6}{\micro\gram\per\milli\liter} of \textsuperscript{233}U in \SI{0.1}{\mole\per\liter} HNO$_3$ (Merck). Circular titanium foils (\SI{99.6}{\percent}, \SI{34}{\milli\meter} in diameter, \SI{25}{\micro\meter} thickness) were used as substrates. For cleaning of the titanium foils, distilled water, \SI{6}{\mole\per\liter} HCl (Merck, \SI{37}{\percent}), isopropanol, and acetone (Merck, \SI{99.5}{\percent}) were used.

\subsubsection{Instrumentation and methods}

A DoD printer setup as described in \cite{Haas2017} was used for printing. Tips with an inner diameter of \SI{200}{\micro\meter} were used. Alpha spectra were recorded with a \SI{2000}{\milli\meter\squared} PIPS detector (Canberra PD2000-40-300AM) in \SI{5}{\milli\meter} distance for yield determination and a \SI{25}{\milli\meter\squared} PIPS detector (Ortec ULTRA\textsuperscript{TM}, \SI{11}{\kilo\electronvolt} FWHM intrinsic resolution) in \SI{100}{\milli\meter} distance was used for recording high resolution spectra.

\subsubsection{Source fabrication and characterization}

\begin{figure*}
	\centering
	\includegraphics[width=1.0\linewidth]{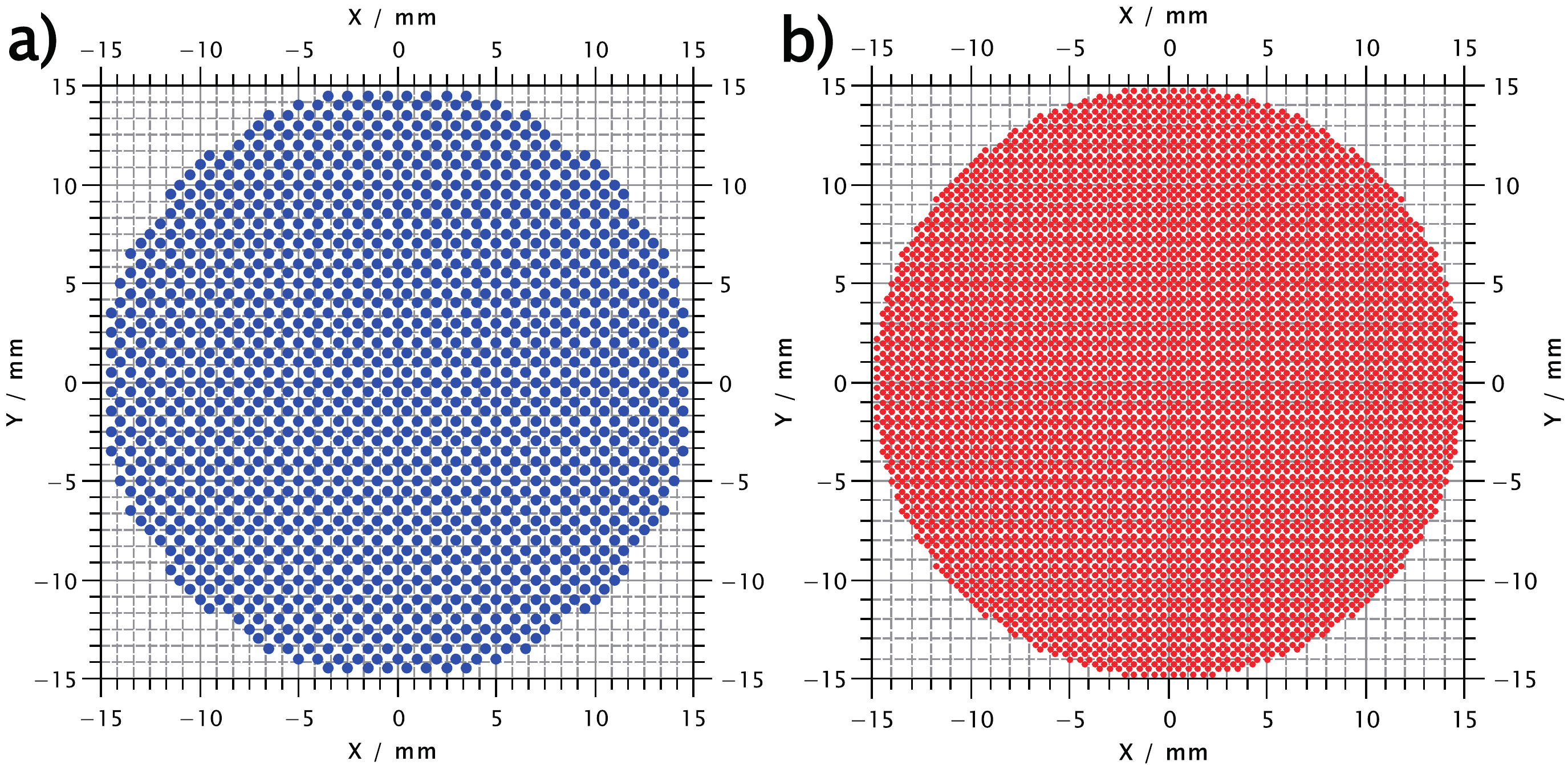}
	\caption{Printing sequences used for source fabrication. Panel a: Sequence containing 1413 coordinates. Panel b: Sequence containing 5637 coordinates. The distances between two coordinates are \SI{0.7}{\milli\meter} (a) and \SI{0.35}{\milli\meter} (b), respectively.}
	\label{fig:2}
\end{figure*}

Five sources were fabricated with a printing sequence containing 1413 drops within a circular area of \SI{30}{\milli\meter} in diameter (see Fig.~\ref{fig:2}~(a)). For this, the \SI{74.2}{\micro\gram\per\milli\liter} solution and a drop volume of \SI{5}{\nano\liter} at a stroke velocity of \SI{100}{\micro\meter\per\milli\second} were used. Two additional sources were fabricated with a printing sequence containing 5637 drops within a circular area of \SI{30}{\milli\meter} in diameter (see Fig.~\ref{fig:2}~(b)). Drop volume and stroke velocity were kept the same at \SI{5}{\nano\liter} and \SI{100}{\micro\meter\per\milli\second}, but the solutions with concentrations of \SI{18.6}{\micro\gram\per\milli\liter} and \SI{138}{\micro\gram\per\milli\liter} were used for these latter two sources. The sources were dried at air in the fume hood after fabrication. Alpha spectra, SEM pictures and RI were processed as described in section~\ref{MPcharacterization}.

\subsection{Chelation by sulfonic acid groups}\label{FS}

\subsubsection{Reagents and materials}

An aliquot of the stock solution was diluted to give a solution with a concentration of \SI{38.9}{\micro\gram\per\milli\liter} of \textsuperscript{233}U in \SI{0.1}{\mole\per\liter} HNO$_3$ (Merck). Pieces of silicon wafers (\SI{300}{\micro\meter} thickness) in various forms with areas of about \SIrange{6.4}{10}{\milli\meter\squared}, functionalized with sulfonic acid groups according to \cite{Krupp2018}, were used as substrates.

\subsubsection{Instrumentation and methods}

Alpha spectra were recorded with a \SI{2000}{\milli\meter\squared} PIPS detector (Canberra PD2000-40-300AM) in \SI{5}{\milli\meter} distance for yield determination and a \SI{25}{\milli\meter\squared} PIPS detector (Ortec ULTRA\textsuperscript{TM}, \SI{11}{\kilo\electronvolt} FWHM intrinsic resolution) in \SI{100}{\milli\meter} distance was used for recording high resolution spectra.

\subsubsection{Source fabrication and characterization}

The functionalized wafers were placed in a crystallizing basin and covered with \SI{10}{\milli\liter} of the \textsuperscript{233}U solution. After \SI{5}{\minute}, \SI{15}{\minute}, \SI{30}{\minute}, \SI{60}{\minute} and \SI{120}{\minute} the wafers were carefully removed from the solution, rinsed with distilled water and dried under an IR lamp. Alpha spectra, SEM pictures and RI were processed as described in section~\ref{MPcharacterization}.

\subsection{Self-adsorption}\label{SA}

\subsubsection{Reagents and materials}

An aliquot of the stock solution was diluted with distilled water to give a concentration of \SI{5.89}{\micro\gram\per\milli\liter} of \textsuperscript{233}U. A titanium foil (\SI{99.6}{\percent}, \SI{25}{\micro\meter} thickness) was used for preparation of the substrates. \SI{0.1}{\mole\per\liter} HNO$_3$ (Merck) was used for the adjustment of different pH values in the aqueous solutions. For cleaning of the titanium foils, distilled water, \SI{6}{\mole\per\liter} HCl (Merck, \SI{37}{\percent}), isopropanol and acetone (Merck, \SI{99.5}{\percent}) were used.

\subsubsection{Instrumentation and methods}

A muffle furnace (Nabertherm, LE 1/11/R7, max. \SI{1100}{\degreeCelsius}) was used for thermal oxidation of the titanium foils at air. Alpha spectra were recorded with a \SI{450}{\milli\meter\squared} PIPS detector (Ortec, CR-SNA-450-100) in \SI{5}{\milli\meter} distance for yield determination and a \SI{25}{\milli\meter\squared} PIPS detector (Ortec ULTRA\textsuperscript{TM}, CU-011-025-300, \SI{11}{\kilo\electronvolt} FWHM intrinsic resolution) in \SI{100}{\milli\meter} distance was used for recording high resolution spectra.

\subsubsection{Source fabrication and characterization}

In a first experimental series, solutions with different pH-values of 2 to 7 were generated as mixtures of \SI{0.1}{\mole\per\liter} HNO$_3$ with distilled water. Pieces with dimensions of \SI{10}{\milli\meter} x \SI{10}{\milli\meter} were pretreated in the muffle furnace at \SI{500}{\degreeCelsius} for one hour. The titanium pieces were put upright into small polyethylene flasks, containing a mixture consisting of \SI{250}{\micro\liter} of the \textsuperscript{233}U solution and \SI{1}{\milli\liter} of the pH-adjusted solution, for one day. Afterwards, the titanium pieces were rinsed with distilled water and dried under an IR lamp in the fume hood. Yield determination was performed by alpha analysis of both sides of the titanium pieces as described in section~\ref{MPcharacterization}.\\
In a second experimental series, Ti pieces with dimensions of \SI{10}{\milli\meter} x \SI{10}{\milli\meter} were thermally oxidized in the muffle furnace at different temperatures between \SIrange{100}{900}{\degreeCelsius} for one hour. The pieces were processed as described before in a solution with a pH value of 5. Yield determination was performed by alpha analysis of both sides of the titanium pieces as described in section~\ref{MPcharacterization}. Three titanium foils, pretreated at temperatures of \SI{20}{\degreeCelsius}, \SI{450}{\degreeCelsius}, and \SI{700}{\degreeCelsius}, were investigated by AFM.\\
At last, two circular titanium foils with a diameter of \SI{30}{\milli\meter} were processed at \SI{450}{\degreeCelsius} and kept in \textsuperscript{233}U solution at a pH value of 5. Another two titanium foils with the same dimensions were sand blasted in advance and processed exactly like the two other foils. Alpha spectra were recorded of one side of those foils. SEM pictures and RI were performed as described in section~\ref{MPcharacterization}.

\subsection{Investigation of the Th daughter recoil efficiency}\label{Eff}

\subsubsection{Reagents and materials}

A stock solution with \SI{50}{\kilo\becquerel} of \textsuperscript{232}U in \SI{6.5}{\milli\liter} \SI{0.1}{\mole\per\liter} HNO$_3$ (Merck) was evaporated under an IR lamp and dissolved again in \SI{8}{\mole\per\liter} HCl (Merck). This process was repeated three times. A column (\SI{50}{\milli\meter} $\times$ \SI{3}{\milli\meter} with Dowex AG 1x8 anion exchanger) was heated up to \SI{55}{\degreeCelsius} and washed several times with \SI{8}{\mole\per\liter} HCl. The \textsuperscript{232}U solution was given onto the column and the daughter nuclides were eluted by 4 $\times$ \SI{2}{\milli\liter} \SI{8}{\mole\per\liter} HCl. The \textsuperscript{232}U was stripped with 5 $\times$ \SI{2}{\milli\liter} \SI{0.5}{\mole\per\liter} HCl. The \textsuperscript{232}U-containing eluate was evaporated to dryness again and the residue was dissolved in \SI{1}{\milli\liter} HNO$_3$ solution at pH 5. Circular titanium foils (Goodfellow, \SI{99.6}{\percent}, \SI{25}{\micro\meter} thickness, \SI{25}{\milli\meter} in diameter) were used as substrates. For cleaning of the titanium foils, distilled water, \SI{6}{\mole\per\liter} HCl (Merck, \SI{37}{\percent}), isopropanol and acetone (Merck, \SI{99.5}{\percent}) were used. A mixture of \SI{10}{\percent} isopropanol (Merck, \SI{99.8}{\percent}) and \SI{90}{\percent} isobutanol (Merck, \SI{99}{\percent}) was used as electrolyte for MP in the plating cells. A 1-mm thick palladium wire (>\SI{99.98}{\percent}) was used as anode for MP.

\subsubsection{Instrumentation and methods}

For depositions (SA and MP) on a limited area with \SI{8}{\milli\meter} in diameter, a vertical cell design \cite{Trautmann1989} was used. For MP, the cell was used in combination with a HV power supply (FUG MCP 350-1250, \SIrange{0}{1250}{\volt}, \SIrange{0}{250}{\milli\ampere}). Alpha spectra were recorded with a \SI{450}{\milli\meter\squared} PIPS detector (Ortec CR-SNA-450-100) in \SI{8}{\milli\meter} distance for yield determination and \SI{450}{\milli\meter\squared} PIPS detectors (Canberra A450-18AM) in \SI{50}{\milli\meter} distance inside an Alpha Analyst system (Canberra, Dual Alpha Spectrometer Upgrade Module 7200) were used for recording high resolution spectra with 4k channels.

\subsubsection{Source fabrication and characterization}

\paragraph*{Self-adsorption}

For the preparation of an SA source, a titanium foil was thermally oxidized in a muffle furnace for \SI{1}{\hour}. Afterwards, it had a copper-colored surface. The oxidized titanium foil was fixed in the plating cell without a spacer disk. The \textsuperscript{232}U solution was filled into the plating cell and Parafilm\textsuperscript{\textcopyright} was used to prevent evaporation. After \SI{24}{\hour}, the solution was carefully removed with a pipette and the titanium foil was rinsed with distilled water and dried under an IR lamp.

\paragraph*{Molecular Plating}

For the preparation of the MP source, the residual solution was evaporated to dryness again and the residue was dissolved in \SI{40}{\micro\liter} \SI{0.1}{\mole\per\liter} HNO$_3$. The solution was mixed with \SI{1}{\milli\liter} isopropanol and \SI{9}{\milli\liter} isobutanol using a vortex mixer. A titanium foil was etched with \SI{6}{\mole\per\liter} HCl and cleaned with distilled water, isopropanol and acetone. Both the titanium foil and the electrolyte solution were given into a plating cell and MP was performed at constant current of about \SI{0.75}{\milli\ampere\per\centi\meter\squared} for \SI{2}{\hour}. After deposition, the source was carefully dried under an IR lamp in a fume hood. Alpha spectra for yield determination were taken of both sources with peak areas of at least \SI{E4}{\counts} to reduce statistical counting errors below \SI{1}{\percent}.

\subsubsection{Recoil collection and investigation of the recoil efficiency}

Both sources were kept in \SI{5}{\milli\meter} distance to a titanium foil (\SI{25}{\milli\meter} in diameter) under vacuum at \SI{8E-3}{\milli\bar}. Monte-Carlo simulations performed with AASI gave \SI[separate-uncertainty = true]{30.9(4)}{\percent} of geometric efficiency for this setup. The collection times were \SI{26}{\day} for the SA source and \SI{5.16}{\day} for the MP source. After the collection, alpha spectra for yield determination were measured of both catcher foils with peak areas of at least \SI{E4}{\counts} to reduce statistical counting errors below \SI{1}{\percent}. Furthermore, qualitative alpha spectra were taken of both sources and the corresponding catcher foils.

\section{Experimental results}\label{ExpRes}

\subsection{Molecular Plating}

\begin{figure}[!htb]
	\centering
	\includegraphics[width=1.0\linewidth]{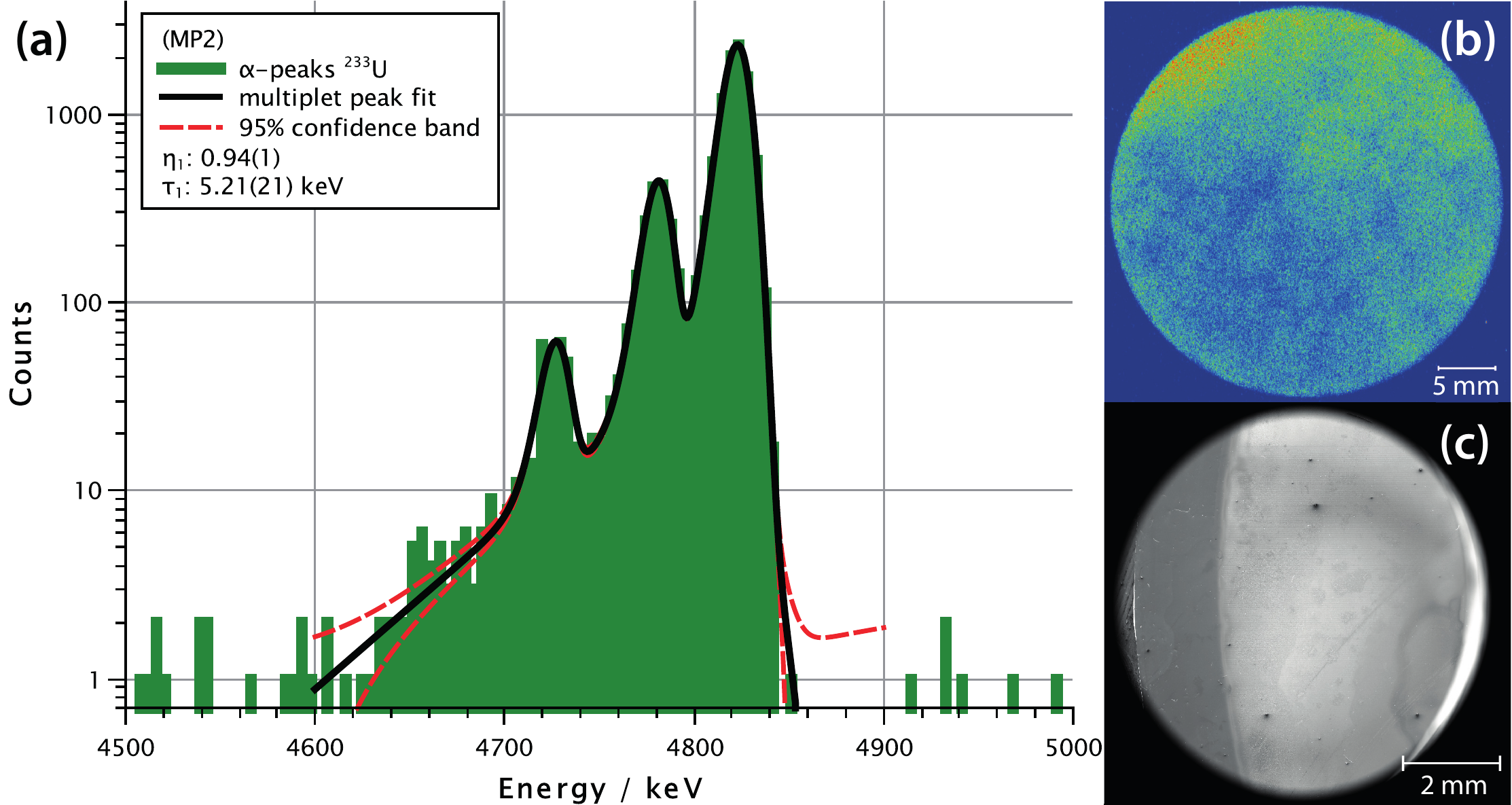}
	\caption{Qualitative alpha spectrum including a multiplet peak fit with confidence band (a), RI (b) and SEM picture (c) of a molecular plated \textsuperscript{233}U recoil ion source (MP2). The source was fabricated on a titanium-coated silicon wafer from a mixture of \SI{10}{\percent} isopropanol and \SI{90}{\percent} isobutanol. Irradiation time for the RI was about \SI{2}{\hour}.}
	\label{fig:3}
\end{figure}

The substrates, plating durations, deposit areas and quantitatively determined areal density of the MP sources are given in Table~\ref{tab5}. The deposition area had a shiny surface. No material layer was observable with a light microscope. The qualitative alpha spectrum of source MP2 is shown in Fig.~\ref{fig:3} together with a RI and a SEM picture. The three peaks of the \textsuperscript{233}U multiplet are well resolved. The RI shows some inhomogeneities of the areal activity distribution. The surface of both, the coated wafer and the deposition area, look very smooth in the SEM picture except for some residues of the solvent.

\begin{table*}[!htb]
	\caption{Parameters and yield of the MP-produced \textsuperscript{233}U recoil ion sources. N,N'-Dimethyl\-formamide (DMF) and mixtures of \SI{90}{\percent} isopropanol and \SI{10}{\percent} isobutanol were used as electrolytes. The areal density refers to \textsuperscript{233}U atoms.}
	\label{tab5}
	\centering
	\begin{tabulary}{1.0\textwidth}{ C | C | @{}C@{} | C | C | @{}C@{} | C }
		\hline
		\mbox{sample} & \mbox{conc.} / \SI{}{\micro\gram\per\milli\liter} & substrate & \mbox{electrolyte} & duration / h & area / \SI{}{\centi\meter\squared} & areal density / \SI{E14}{\per\centi\meter\squared} \\ 
		\hline
		MP1 & 5.25 & wafer & DMF & 1.0 & 7.1 & 0.69(1) \\ 
		MP2 & 34.4 & wafer & IP/IB & 1.5 & 7.1 & 3.82(4) \\ 
		MP3 & 21.5 & Ti foil & IP/IB & 1.0 & 0.5 & 2.48(3) \\ 
		MP4 & 21.5 & Ti foil & IP/IB & 1.5 & 0.5 & 19.1(2) \\ 
		MP5 & 21.5 & Ti foil & IP/IB & 2.0 & 0.5 & 19.8(2) \\ 
		\hline 
	\end{tabulary}
\end{table*} 

\subsection{Drop-on-Demand inkjet printing}

\begin{figure}[!htbp]
	\centering
	\includegraphics[width=1.0\linewidth]{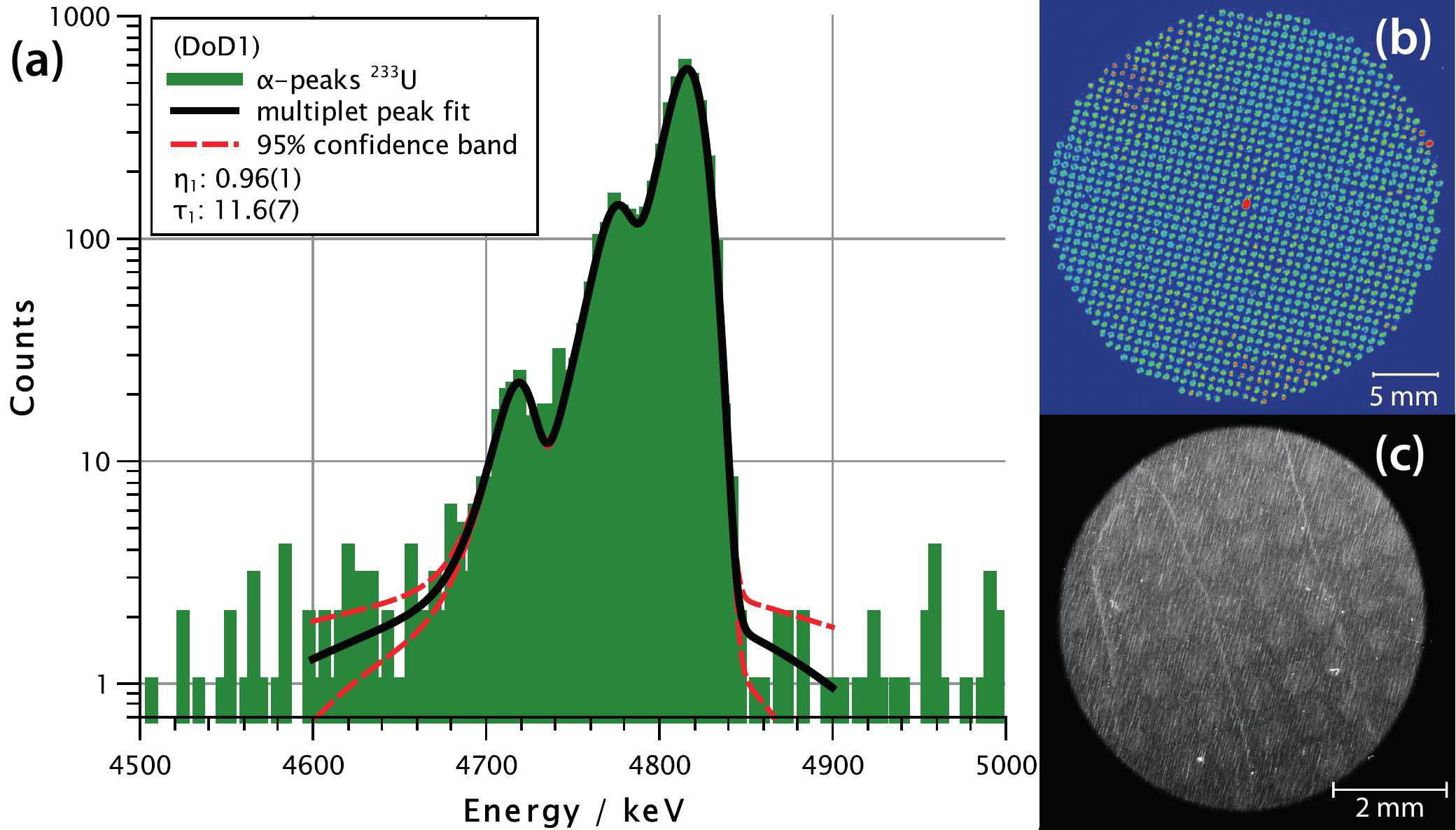}
	\caption{Qualitative alpha spectrum (a) of a DoD-printed \textsuperscript{233}U recoil ion source (DoD1) on titanium foil as well as RI (b) and SEM picture (c). Irradiation time for the RI was about \SI{2}{\hour}.}
	\label{fig:4}
\end{figure}

The number of drops, drop volume, printed area and yield of the DoD sources are given in Table~\ref{tab6}. The droplets dried within seconds during the printing process. No residues were observable with a light microscope. The qualitative alpha spectrum of the printed source DoD1 containing 1413 drops as well as its RI and SEM picture are given in Fig.~\ref{fig:4}. The three peaks are resolved. The RI shows a macroscopically homogeneous areal distribution of the activity. The size of the deposits varies within a range of about \SI{200}{\micro\meter} due to surface effects of the titanium foil. The deposits of the single drops are faintly visible in the SEM picture thanks to a high height contrast by the rough surface of the titanium foil.

\begin{table*}[!htbp]
	\caption{Parameters for DoD printing and yield of the \textsuperscript{233}U recoil ion sources. The areal density refers to the \textsuperscript{233}U atoms on the area given in the table.}
	\label{tab6}
	\centering
	\begin{tabulary}{1.0\textwidth}{ C | @{}C@{} | C | C | C | @{}C@{} }
		\hline 
		\mbox{sample} & \mbox{conc.} / \SI{}{\micro\gram\per\milli\liter} & drops & drop volume / \SI{}{\nano\liter} & area / \SI{}{\centi\meter\squared} & areal density / \SI{E14}{\per\centi\meter\squared} \\ 
		\hline
		DoD1 & 74.2 & 1413 & 5.0 & 7.1 & 4.45(4) \\ 
		DoD2 & 74.2 & 1413 & 5.0 & 7.1 & 4.60(5) \\ 
		DoD3 & 74.2 & 1413 & 5.0 & 7.1 & 1.83(2) \\ 
		DoD4 & 74.2 & 1413 & 5.0 & 7.1 & 1.81(2) \\ 
		DoD5 & 74.2 & 1413 & 5.0 & 7.1 & 1.82(2) \\ 
		DoD6 & 18.6 & 5637 & 5.0 & 7.1 & 1.23(1) \\ 
		DoD7 & 138 & 5637 & 5.0 & 7.1 & 6.84(7) \\ 
		\hline 
	\end{tabulary}
\end{table*} 

\subsection{Chelation by sulfonic acid groups}

\begin{figure}[!htbp]
	\centering
	\includegraphics[width=1.0\linewidth]{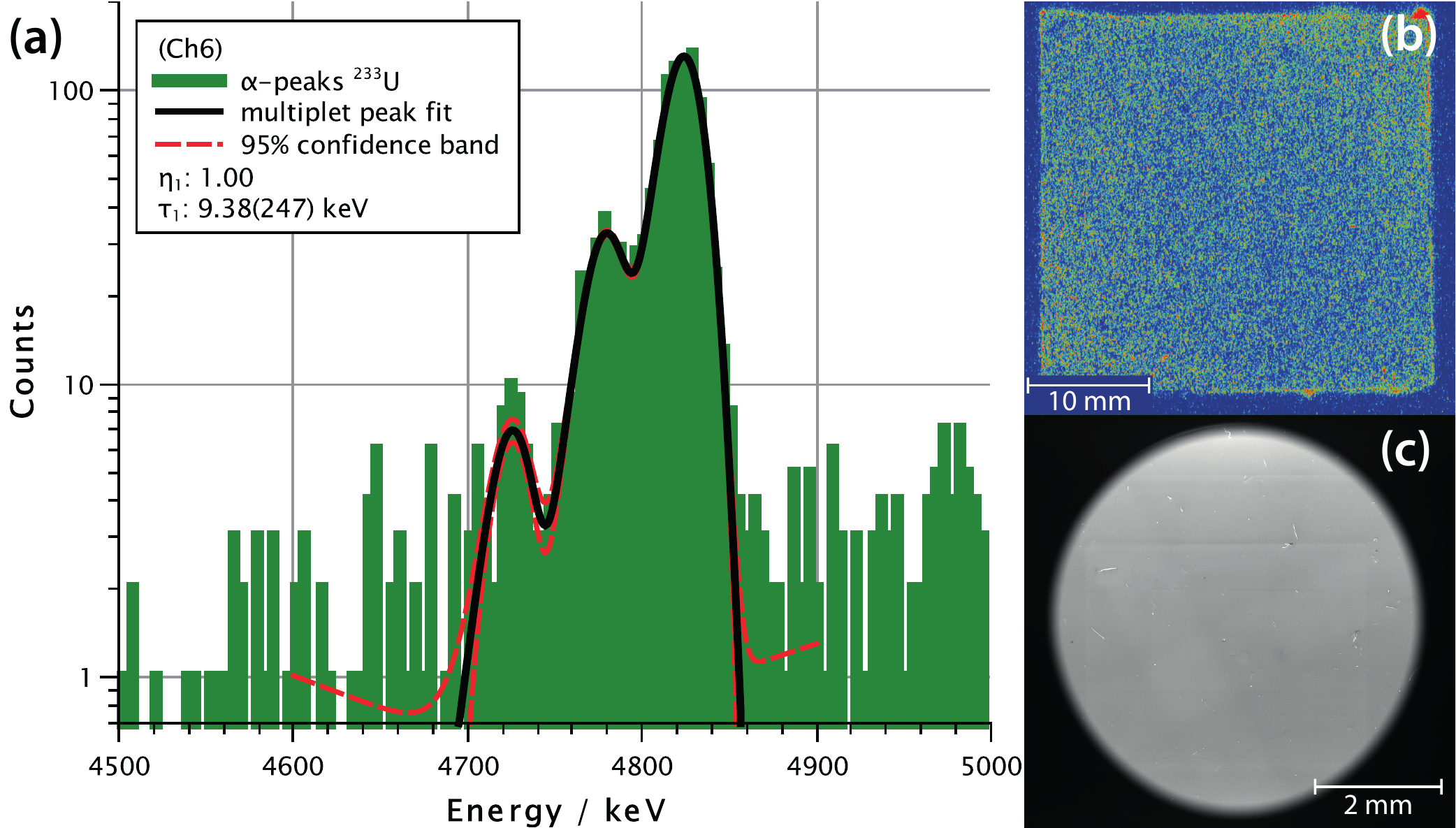}
	\caption{Qualitative alpha spectrum (a) of a \textsuperscript{233}U recoil ion source (Ch6) fabricated by chelation on a functionalized silicon wafer piece as well as RI (b) and SEM picture (c). Irradiation time for the RI was about \SI{1}{\hour}.}
	\label{fig:5}
\end{figure}

The contact times, areas of the wafer pieces and areal densities of the chelated \textsuperscript{233}U deduced from the quantitative alpha spectra are given in Table~\ref{tab7}. No visible change occurred on the surface of the wafer pieces due to the fabrication process. The qualitative alpha spectrum of the source Ch6 including the multiplet peak fit is given in Fig.~\ref{fig:5}. Due to the very low areal density on the wafers, the count rate is very low. The three peaks are well-resolved and very narrow. The RI of the same source in Fig.~\ref{fig:5} shows a homogeneous, yet not very dense areal distribution of the activity.

\begin{table*}[!htbp]
	\caption{Parameters for chelation by sulfonic acid groups on silicon wafer pieces and areal densities. Contact time with the radioactive solution and areas of the silicon wafers are given. The areal density refers to \textsuperscript{233}U atoms.}
	\label{tab7}
	\centering
	\begin{tabulary}{1.0\textwidth}{ C | C | C | C }
		\hline 
		\mbox{sample} & contact time / min & area / \SI{}{\centi\meter\squared} & areal density / \SI{E14}{\per\centi\meter\squared} \\ 
		\hline
		Ch1 & 5 & 7.40 & 0.210(2) \\ 
		Ch2 & 15 & 9.97 & 0.122(1) \\ 
		Ch3 & 30 & 7.98 & 0.149(1) \\ 
		Ch4 & 30 & 8.74 & 0.109(1) \\ 
		Ch5 & 60 & 6.48 & 0.235(2) \\ 
		Ch6 & 120 & 9.89 & 0.238(2) \\ 
		\hline 
	\end{tabulary}
\end{table*} 

\subsection{Self-adsorption}

\begin{figure*}[!htb]
	\centering
	\includegraphics[width=1.0\linewidth]{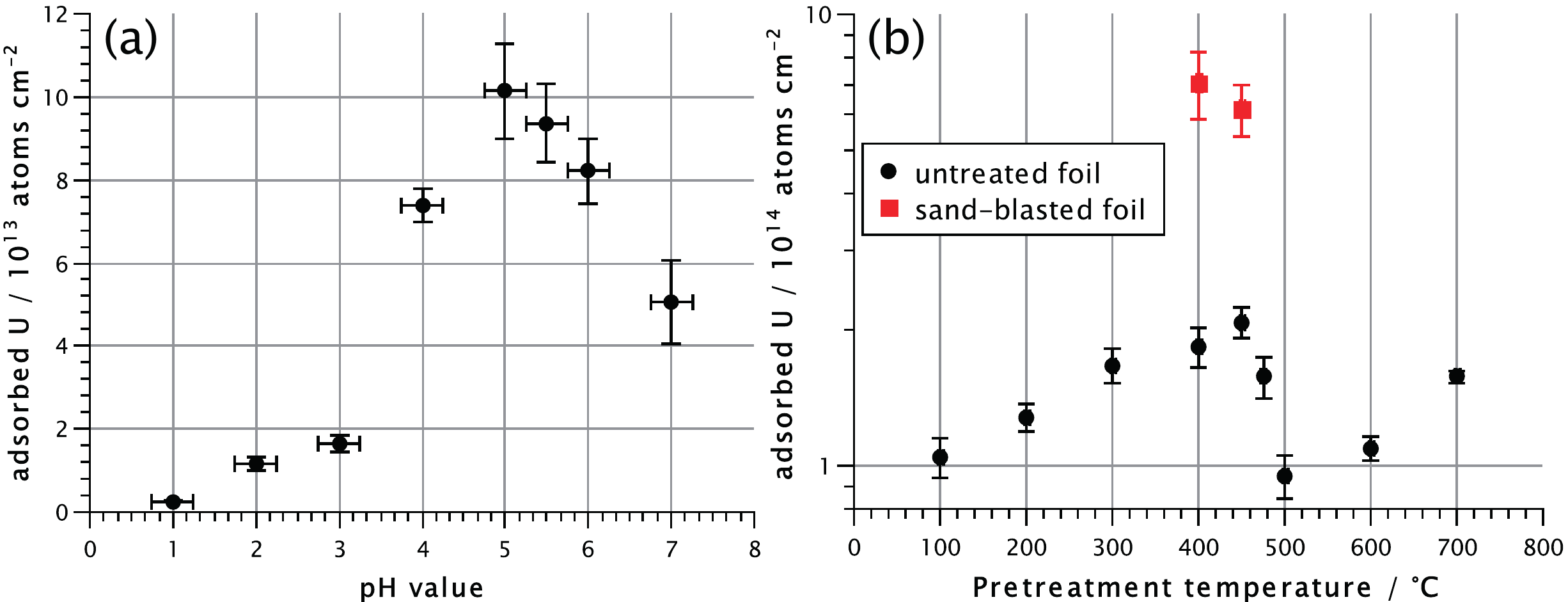}
	\caption{(a) Areal density of \textsuperscript{233}U atoms on Ti foil thermally oxidized at \SI{500}{\degreeCelsius} for \SI{1}{\hour} as a function of pH value. (b) Areal density of \textsuperscript{233}U atoms obtained at pH 5 on Ti foils as a function of the oxidation temperature applied before the adsorption exposure.}
	\label{fig:6}
\end{figure*}

Fig.~\ref{fig:6} shows the \textsuperscript{233}U density of the SA sources as functions of the pH value of the solution (panel a) and the pretreatment temperature (panel b). The maximum yield was reached from a solution with a pH value of about 5, resulting in about \SI{40}{\nano\gram\per\centi\meter\squared} (\SI{E14}{\atoms\per\centi\meter\squared}) on thermally oxidized titanium foils at \SI{500}{\degreeCelsius}. This was increased to about \SI{80}{\nano\gram\per\centi\meter\squared} (\SI{2E14}{\atoms\per\centi\meter\squared}) by changing the pretreatment temperature of the titanium foils to \SI{450}{\degreeCelsius}. Sand-blasted titanium foils showed \SI{270}{\nano\gram\per\centi\meter\squared} (\SI{7E14}{\atoms\per\centi\meter\squared}) of \textsuperscript{233}U when prepared under the same conditions. To inspect a potentially big influence of different surface roughnesses, the corresponding values are listed in Table~\ref{tab8}.

\begin{table}[!htb]
	\caption{Root mean square (RMS) roughness of titanium foils pretreated at different temperatures for \SI{1}{\hour}. Atomic force microscopy was used for the investigation of the RMS roughness.}
	\label{tab8}
	\centering
	\begin{tabulary}{1.0\linewidth}{ C | C }
		\hline
		pretreatment temperature / \SI{}{\degreeCelsius} & RMS roughness / \SI{}{\nano\meter} \\ 
		\hline
		20 & 39(2) \\ 
		450 & 24(10) \\ 
		700 & 40(10) \\ 
		\hline 
	\end{tabulary}
\end{table} 

\begin{figure}[!htb]
	\centering
	\includegraphics[width=1.0\linewidth]{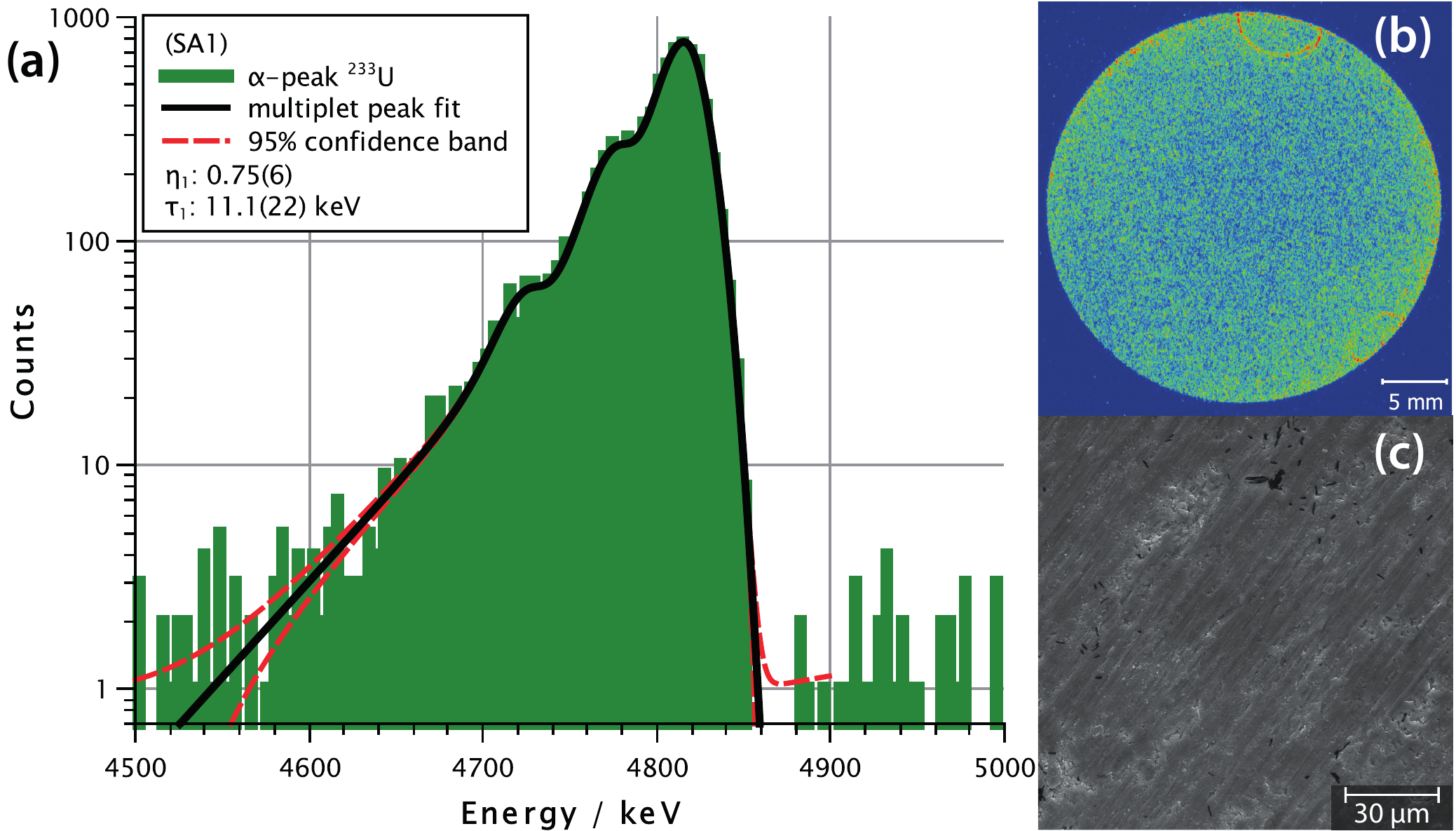}
	\caption{Qualitative alpha spectrum (a) as well as RI (b) and SEM picture (c) of a \textsuperscript{233}U recoil ion source (SA1) fabricated by self-adsorption on a smooth titanium foil thermally oxidized at \SI{450}{\degreeCelsius} for \SI{1}{\hour}. Irradiation time for the RI was about \SI{2}{\hour}.}
	\label{fig:7}
\end{figure}

\begin{figure}[!htb]
	\centering
	\includegraphics[width=1.0\linewidth]{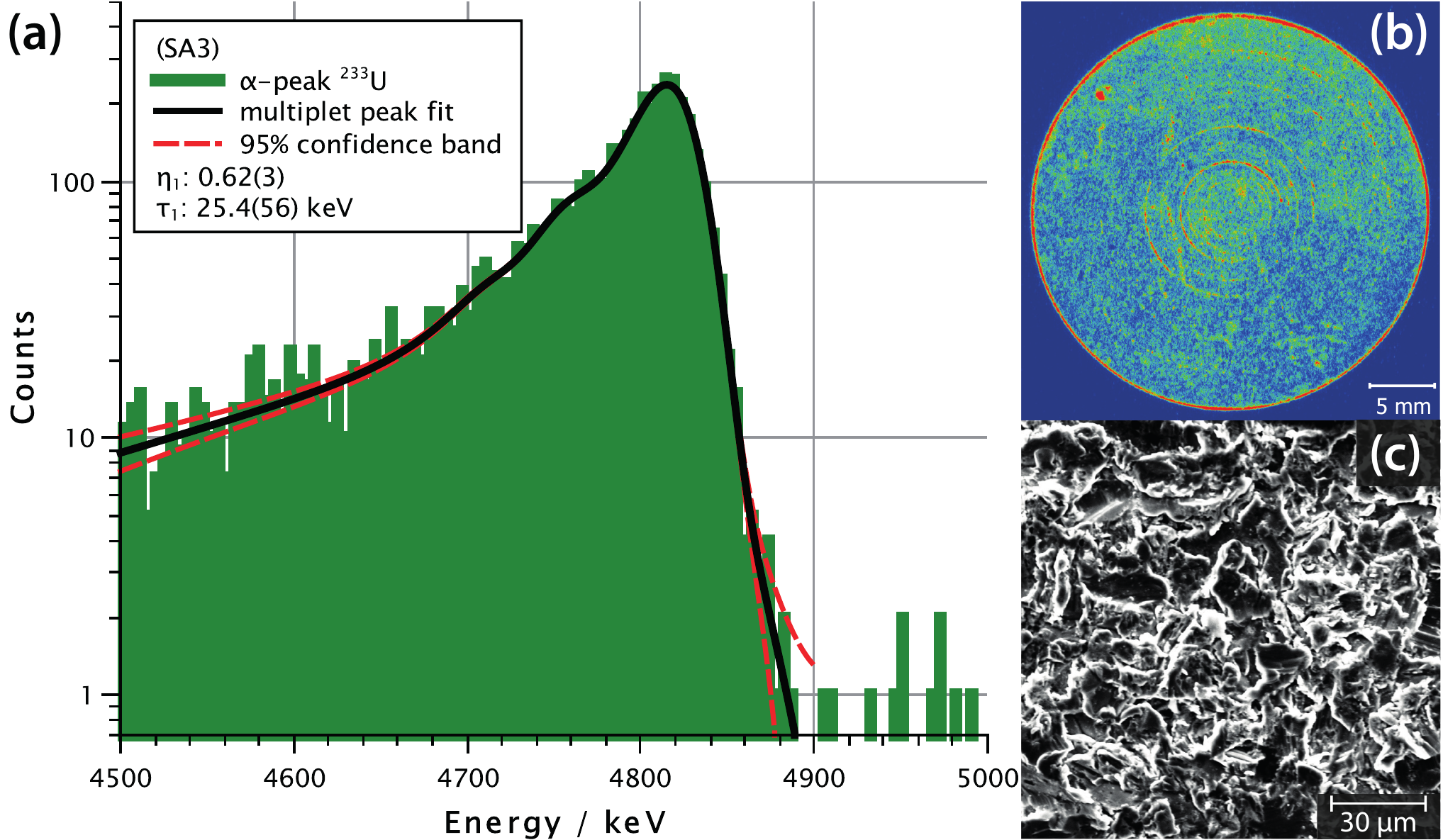}
	\caption{Qualitative alpha spectrum (a) as well as RI (b) and SEM picture (c) of a \textsuperscript{233}U recoil ion source (SA3) fabricated by self-adsorption on a sand-blasted titanium foil thermally oxidized at \SI{450}{\degreeCelsius} for \SI{1}{\hour}. Irradiation time for the RI was about \SI{2}{\hour}.}
	\label{fig:8}
\end{figure}

The parameters for fabrication and yields of the SA sources are given in Table~\ref{tab9}. No visible change occurred on the surface of the titanium foils after fabrication. The qualitative alpha spectrum of a source on a non-sand-blasted titanium foil (SA1) including the multiplet peak fit is given in Fig.~\ref{fig:7}~(a). The two peaks at lower energies are visible as shoulders of the peak at \SI{4824}{\kilo\electronvolt} and are not well-resolved. The RI in Fig.~\ref{fig:7}~(b) shows a homogeneous areal distribution of activity. In the SEM picture in Fig.~\ref{fig:7}~(c), the height contrast of the titanium foil dominates, as this has a quite rough surface even without sand-blasting. No signal originating from the \textsuperscript{233}U can be seen. The qualitative alpha spectrum of a source on a sand-blasted titanium foil (SA3) as well as its RI and SEM pictures are given in Fig.~\ref{fig:8}. The left shoulder of the main alpha peak at \SI{4824}{\kilo\electronvolt} is larger and the two peaks at lower energies can barely be identified. RI shows a quite homogeneous areal distribution of the activity but also shows artifacts due to the very rough surface structure of the foil, which is clearly visible in the SEM picture.

\begin{table*}[!htb]
	\caption{Parameters for self-adsorption on preheated titanium foils and yields. The areal density refers to \textsuperscript{233}U atoms.}
	\label{tab9}
	\centering
	\begin{tabulary}{1.0\textwidth}{ C | @{}C@{} | @{}C@{} | C | C@{} }
		\hline
		\mbox{sample} & sand-blasted & pretreatment temperature / \SI{}{\degreeCelsius} & area / \SI{}{\centi\meter\squared} & areal density / \SI{E14}{\per\centi\meter\squared} \\ 
		\hline
		SA1 & no & 450 & 7.1 & 2.33(2) \\ 
		SA2 & no & 450 & 7.1 & 1.90(2) \\ 
		SA3 & yes & 450 & 7.1 & 6.73(7) \\ 
		SA4 & yes & 450 & 7.1 & 1.58(2) \\ 
		\hline 
	\end{tabulary}
\end{table*} 

\subsection{Comparison of fitting results of all four methods}

\begin{figure*}[!htb]
	\centering
	\includegraphics[width=1.0\linewidth]{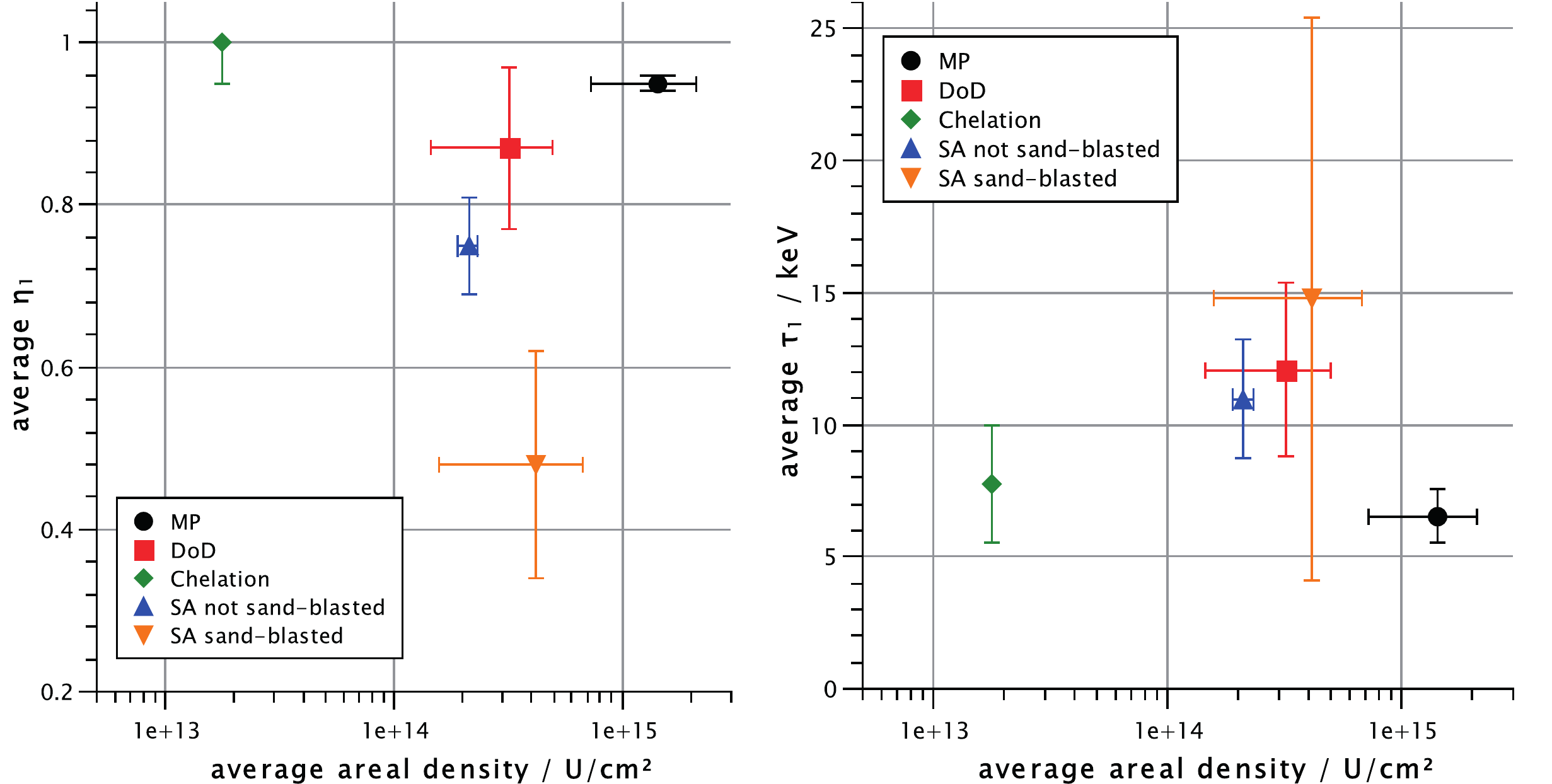}
	\caption{Plotted average fit parameters $\eta_1$ (left) and $\tau_1$ (right) for each fabrication method against the average areal \textsuperscript{233}U density of the sources.}
	\label{fig:9}
\end{figure*}

A complete list including all results of the fit parameters is given in Table~\ref{tab11} in the appendix. The average values of the most significant parameters, $\eta_1$ and $\tau_1$, are plotted against the average areal \textsuperscript{233}U density for each fabrication method in Fig.~\ref{fig:9}. The fit parameter $\sigma_1$ is also significant for the qualitative evaluation, but is identical within error bars for all fabrication methods. Therefore, $\sigma_1$ is not included in the plots. The other parameters, $\sigma_2$ and $\tau_2$, are less significant and have in general much higher values (see Table~\ref{tab11} in the appendix).

\subsection{Recoil efficiency of MP and SA prepared sources}

\begin{figure}[!htb]
	\centering
	\includegraphics[width=0.8\linewidth]{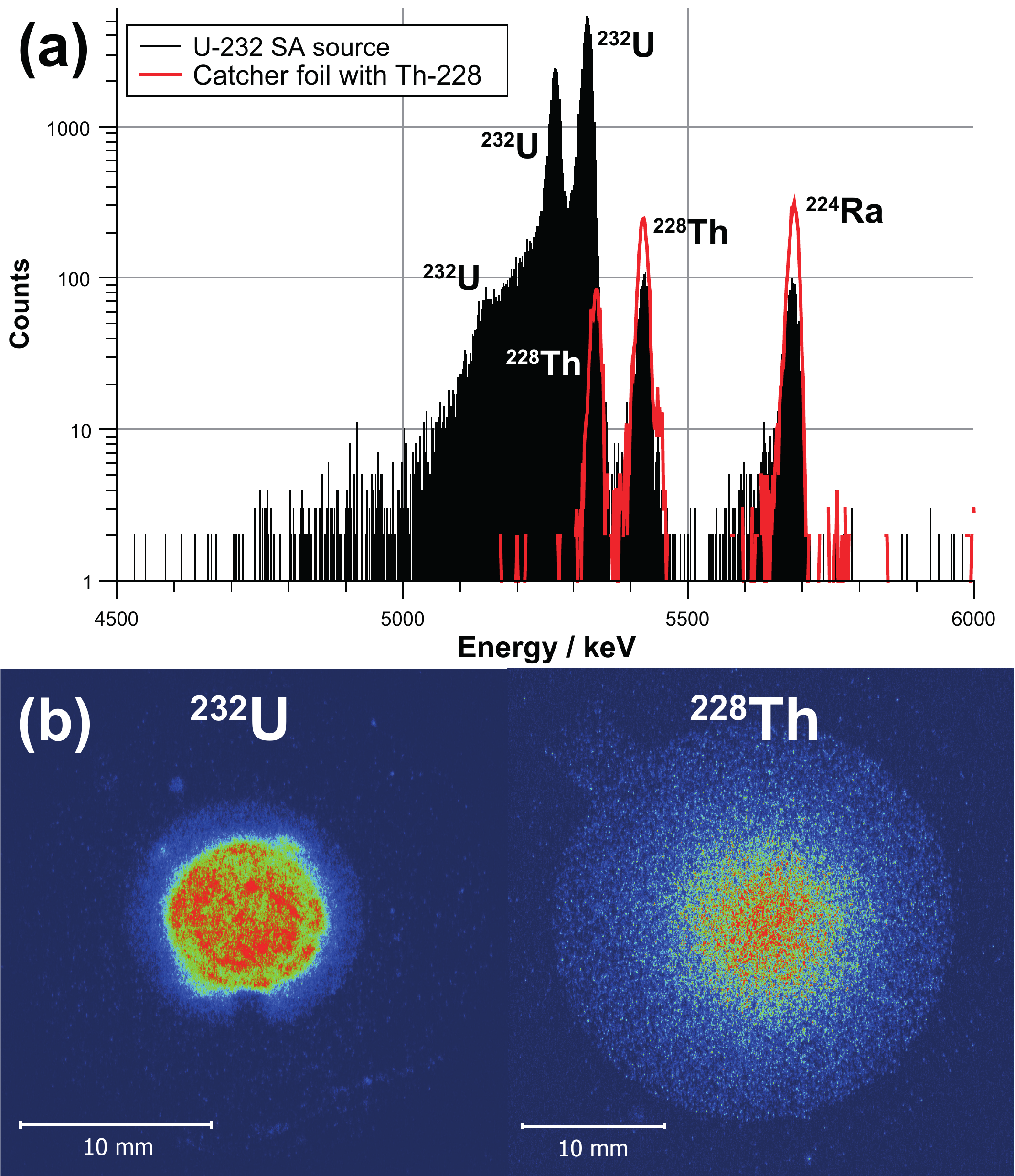}
	\caption{Qualitative alpha spectrum (a) and RI (b) of the \textsuperscript{232}U SA source and the corresponding \textsuperscript{228}Th recoil catcher foil. The spectra were measured for \SI{1}{\hour} in case of the SA source and \SI{5}{\hour} for the corresponding catcher foil, to obtain sufficient counts. The additional line at \SI{5685}{\kilo\electronvolt} belongs to \textsuperscript{224}Ra, the daughter nuclide of \textsuperscript{228}Th. Irradiation time for the RI was about \SI{15}{\minute} for the SA source and \SI{2}{\hour} for the catcher foil.}
	\label{fig:10}
\end{figure}

The active areas of both fabricated sources were visible with the naked eye. The contact area of the SA source turned from copper-colored to violet and the deposition of the MP source had a bright yellow color. In contrast, no visible change occurred on the catcher foils. The geometric detection efficiency of the catcher foils was determined by Monte-Carlo simulations with AASI for the quantitative measurements. The determined \textsuperscript{232}U source activities, the parameters and yields for the \textsuperscript{228}Th recoil collection and the resulting recoil efficiencies are given in Table~\ref{tab10}. The error of the recoil efficiency was calculated by a Gaussian error propagation, using the counting errors of the quantitative measurements as well as the errors of the geometric efficiencies for recoil collection and alpha detection. In Fig.~\ref{fig:10}, the qualitative spectra as well as the RI of the SA source and the corresponding catcher foil are depicted.

\begin{table*}[!htb]
	\caption{\textsuperscript{232}U activities of the SA and MP source and parameters and yields for the \textsuperscript{228}Th recoil collection.}
	\label{tab10}
	\centering
	\begin{tabulary}{1.0\textwidth}{ C | @{}C@{} | @{}C@{} | @{}C@{} | @{}C@{} | C }
		\hline
		\mbox{source} & \textsuperscript{232}U activity / \SI{}{\becquerel} & collection time / \SI{}{\day} & geometric efficiency (AASI) / \SI{}{\percent} & collected \textsuperscript{228}Th / \SI{}{\becquerel} & recoil efficiency / \SI{}{\percent} \\ 
		\hline
		SA & 1300(9) & 26.0 & 16.6(5) & 10.7(1) & 94.2(33) \\ 
		MP & 18800(60) & 5.16 & 16.6(5) & 3.13(2) & 10.5(11) \\ 
		\hline 
	\end{tabulary}
\end{table*} 

\section{Simulation results}\label{ExpSim}

\begin{figure}[!htb]
	\centering
	\includegraphics[width=0.8\linewidth]{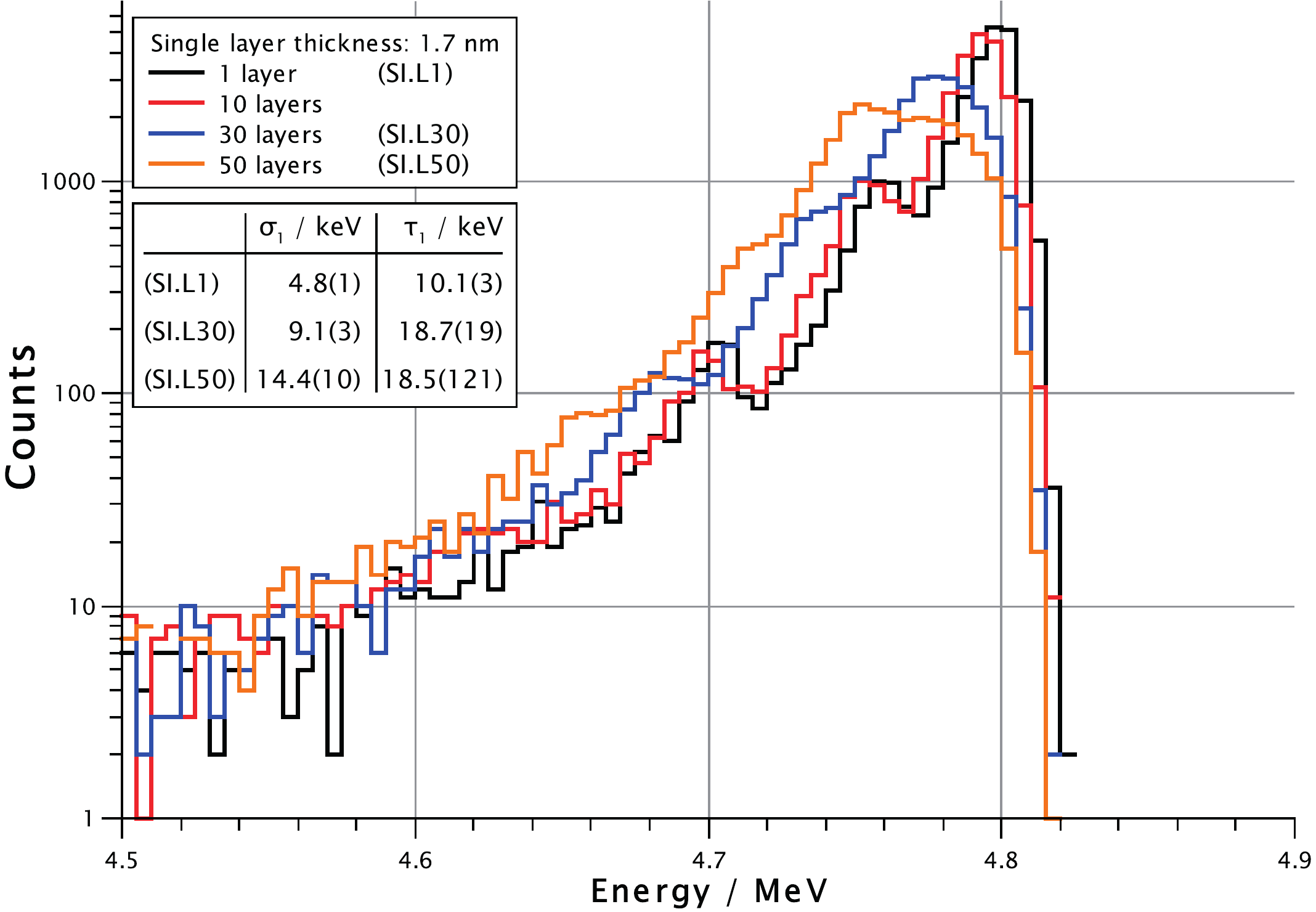}
	\caption{AASI simulations of \textsuperscript{233}U sources with different numbers of atomic layers. The source material was set to UO$_2$(OH)$_2$$\times$H$_2$O with a diameter of \SI{30}{\milli\meter} on a Ti backing with \SI{25}{\micro\meter} thickness. The layer thickness was varied from \SI{1.7}{\nano\meter} (representing a monolayer) up to \SI{85}{\nano\meter}. The source-detector-distance was set to \SI{100}{\milli\meter} and a number of \SI{E6}{} decays were simulated for each parameter variation.}
	\label{fig:11}
\end{figure}

\begin{figure}[!htb]
	\centering
	\includegraphics[width=0.8\linewidth]{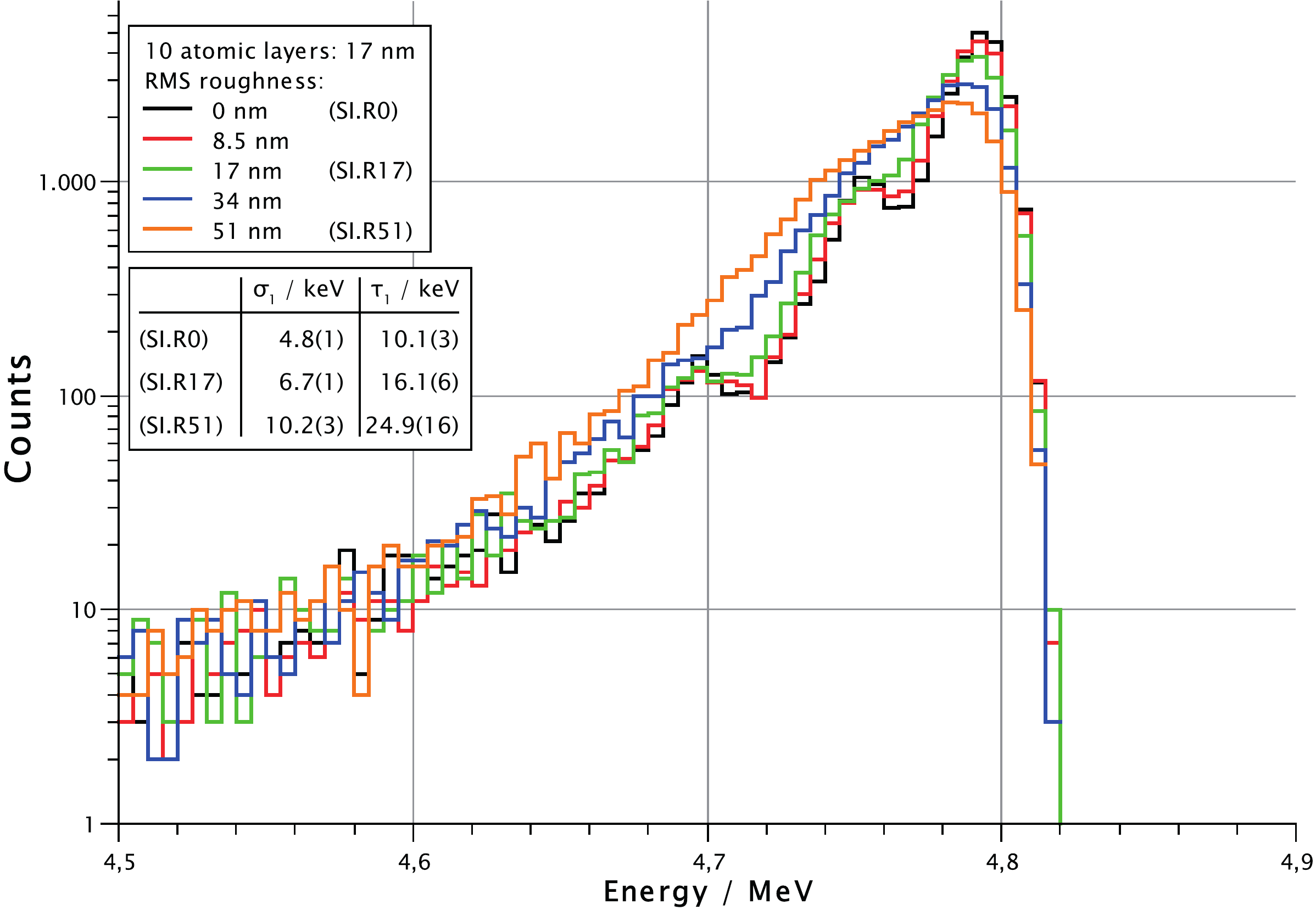}
	\caption{AASI simulations of \textsuperscript{233}U sources with different root mean square roughnesses. The source material was set to UO$_2$(OH)$_2$$\times$H$_2$O with a diameter of \SI{30}{\milli\meter} and a thickness of \SI{17}{\nano\meter} (representing 10 atomic layers) on a Ti backing with \SI{25}{\micro\meter} thickness. The source-detector-distance was set to \SI{100}{\milli\meter} and a number of \SI{E6}{} decays were simulated for each parameter variation. The RMS roughness was varied in the range of \SIrange{0}{51}{\nano\meter}.}
	\label{fig:12}
\end{figure}

\begin{figure}[!htb]
	\centering
	\includegraphics[width=0.8\linewidth]{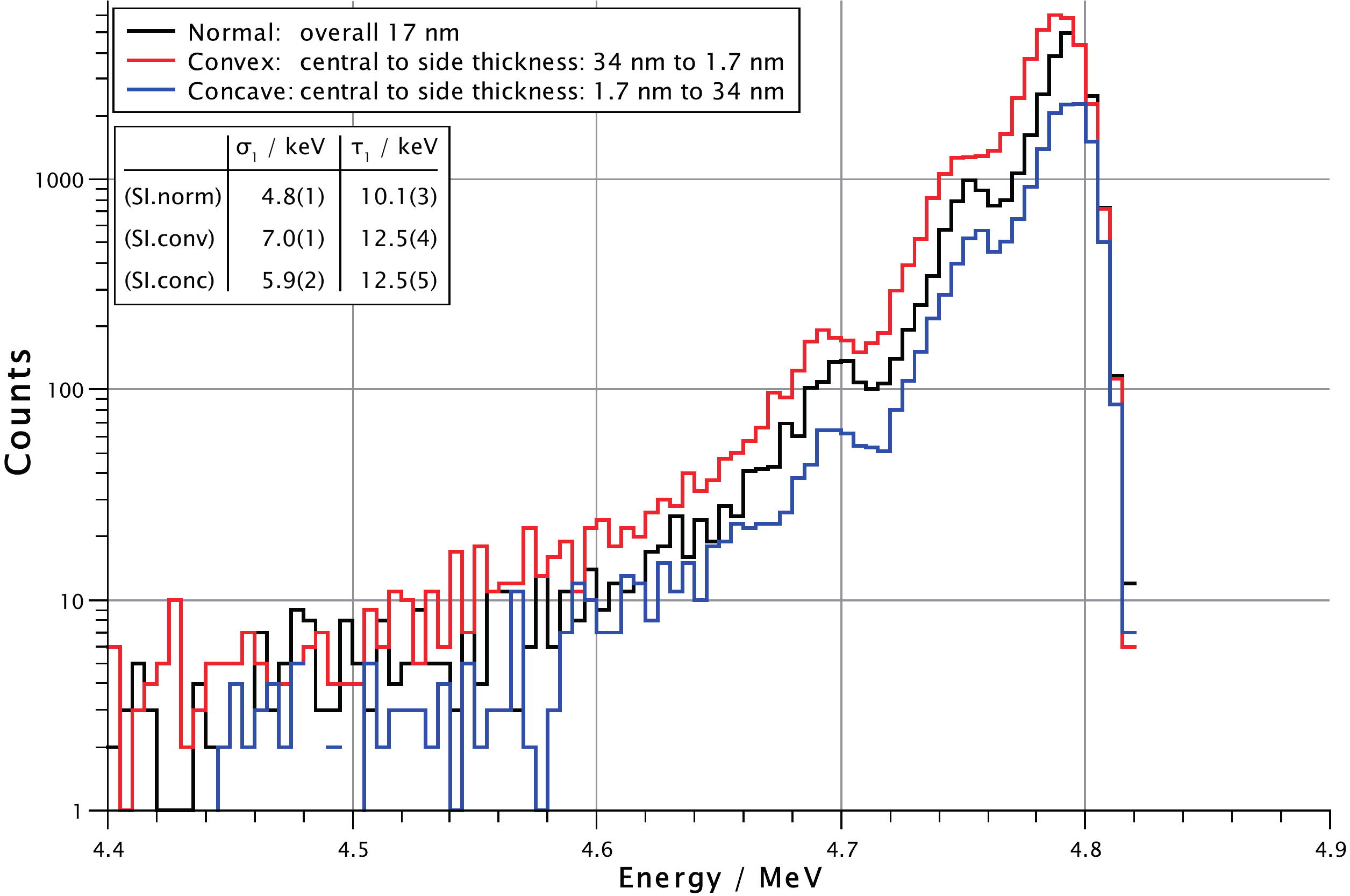}
	\caption{AASI simulations of \textsuperscript{233}U sources with different layer shapes. Normal represents a cylindrical-shape, as the central thickness and the thickness at the side of the source layer are equal. Convex and concave source shapes are described with a paraboloid of revolution, as described in \cite{Siiskonen2005}. The source material was set to UO$_2$(OH)$_2$$\times$H$_2$O with a diameter of \SI{30}{\milli\meter} and a thickness of \SI{17}{\nano\meter} (representing 10 atomic layers) on a Ti backing with \SI{25}{\micro\meter} thickness. The source-detector-distance was set to \SI{100}{\milli\meter} and a number of \SI{E6}{} decays were simulated for each parameter variation.}
	\label{fig:13}
\end{figure}

\begin{figure}[!htb]
	\centering
	\includegraphics[width=1.0\linewidth]{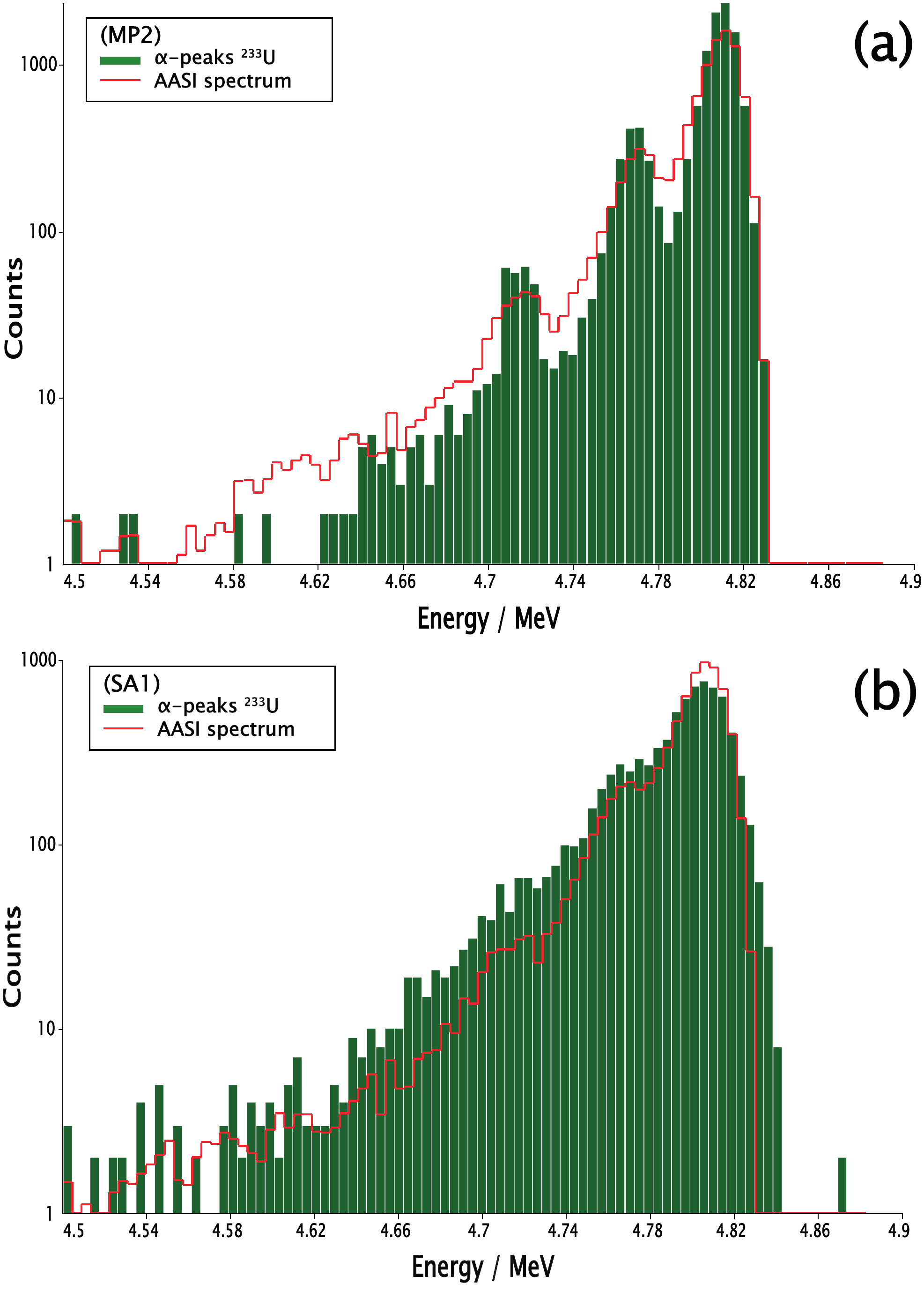}
	\caption{Qualitative comparison of experimental alpha spectra with AASI simulations. For the simulations, the source material was set to UO$_2$(OH)$_2$$\times$H$_2$O with a diameter of \SI{30}{\milli\meter} and a thickness of \SI{17}{\nano\meter} (representing 10 atomic layers) on a Ti backing with \SI{25}{\micro\meter} thickness. The source-detector-distance was set to \SI{100}{\milli\meter} and a number of \SI{E6}{} decays were simulated. In (a), for comparison with the molecular plated source (MP2), the RMS roughness was set to \SI{0}{\nano\meter}, while for comparison with the self-adsorbed source (SA1) in (b) it was set to \SI{34}{\nano\meter}.}
	\label{fig:14}
\end{figure}

As two fabrication methods (MP and DoD) cannot produce specifically monolayers, simulations are necessary to show the influence of several atomic layers on the alpha spectra of \textsuperscript{233}U sources. The effects of different numbers of source layers (Fig.~\ref{fig:11}), different RMS roughnesses (Fig.~\ref{fig:12}) and different macroscopic layer shapes (Fig.~\ref{fig:13}) on simulated alpha spectra are shown. Convex and concave source shapes are described with a paraboloid of revolution, as described in \cite{Siiskonen2005}. The most important fit parameters, $\sigma_1$ and $\tau_1$, are given in a legend for some simulated spectra to show quantitatively the effect of different simulation parameters on the alpha spectrum. The residual fit parameters are given in Table~\ref{tab11}. The influence of different chemical species (see Table~\ref{tab3}) on simulated alpha spectra was also investigated, for sources having 10 atomic layers. The simulated spectra were identical and are therefore not plotted individually. The information on source material, backing and detector, which were used as parameters in AASI, is given in the caption of the figures. The varied parameters are given in the legends of the figures. A constant number of \SI{E6}{} simulated decays and a constant bin size of \SI{5}{\kilo\electronvolt} was used for all simulations. The source-detector-distance was kept constant at \SI{100}{\milli\meter} and the PIPS detector properties of the Ortec ULTRA\textsuperscript{TM}, as employed to obtain the qualitative spectra, were used to enable a direct comparison between experimental and simulated results. For a qualitative comparison of experimental spectra with these simulations, two exemplary alpha spectra of a molecular plated source (MP2) and a self-adsorbed source (SA1) were fitted with simulations in AASI. These are shown in Fig.~\ref{fig:14}.

\section{Discussion}\label{Discu}

\subsection{AASI simulations}

The simulation results of Fig.~\ref{fig:11} show that it is not simple to differentiate between a \textsuperscript{233}U source containing a monolayer and one with 10 atomic layers based on the recorded alpha spectra. The difference can be seen by an energy shift of the peak maximum in the fit evaluation, because the error of the fitted peak position $\mu$ is about \SI{0.5}{\kilo\electronvolt}. Both simulated spectra have the same shape, except for a slight shift of one bin (\SI{5}{\kilo\electronvolt}) to lower energies in case of the alpha source with 10 atomic layers. The systematic uncertainty of the alpha spectra collected with our used alpha-spectrometry setup is estimated to be about \SI{10}{\kilo\electronvolt}; we are therefore not able to extract any information on the number of atomic layers from the peak positions in our experimental spectra. The Monte-Carlo simulations suggest this shift to occur predominantly because of alpha particles originating from inside the source material and less of these occurring from the top-most layers, as inferred from the fact that the number of decays did not increase proportionally to the number of atomic layers. At 20 to 30 atomic layers, the three single peaks are increasingly less well-resolved. At 50 atomic layers, only a single broad peak is visible. This indicates that with an increasing number of atomic layers in the source material, the alpha peak becomes broader, corresponding to an increase of $\sigma_1$ from \SI[separate-uncertainty = true]{4.8(1)}{\kilo\electronvolt} to \SI[separate-uncertainty = true]{14.4(10)}{\kilo\electronvolt}.\\
The simulation results depicted in Fig.~\ref{fig:12} show a significant increase of the tailing with increasing RMS roughness. Also in this case it is not possible to differentiate between a source with a smooth surface and one with a RMS roughness of \SI{8.5}{\nano\meter} with $\tau_1$ at about \SI{10}{\kilo\electronvolt}. The influence only becomes significant at higher RMS roughnesses. At \SI{17}{\nano\meter} RMS roughness, the three single peaks start to merge due to a higher tailing $\tau_1$ of about \SI[separate-uncertainty = true]{16.1(6)}{\kilo\electronvolt}. At \SI{51}{\nano\meter} RMS roughness, the three peaks are not distinguishable any more due to the very high tailing $\tau_1$ of about \SI[separate-uncertainty = true]{24.9(16)}{\kilo\electronvolt}. Additionally, the peak height starts to decrease but the peak area remains constant. The parameter $\sigma_1$ is linked to the increasing tailing, indicated by an increase from \SI[separate-uncertainty = true]{4.8(1)}{\kilo\electronvolt} up to \SI[separate-uncertainty = true]{10.2(3)}{\kilo\electronvolt}.\\
The simulation results of Fig.~\ref{fig:13} pertain to the case of different shapes (convex/concave rotational paraboloid) of the source layer. The simulations show a minor, but clear difference of the alpha peak shape for the three cases. The fit parameters $\eta_1$, $\sigma_1$ and $\tau_1$ are not significantly affected by the convex or concave layer shape. A slight broadening is indicated by an increase of $\sigma_1$ from \SI[separate-uncertainty = true]{4.8(1)}{\kilo\electronvolt} to \SI[separate-uncertainty = true]{7.0(1)}{\kilo\electronvolt} (convex) and \SI[separate-uncertainty = true]{5.9(2)}{\kilo\electronvolt} (concave). Additionally, the tailing parameter $\tau_1$ is increased in both cases from \SI[separate-uncertainty = true]{10.1(3)}{\kilo\electronvolt} to about \SI[separate-uncertainty = true]{12.5(5)}{\kilo\electronvolt}. More alpha particles reach the detector from a convex source while a concave source shows the opposite effect.

\subsection{Molecular Plating}

Thin \textsuperscript{233}U sources with areal densities between \SI{7E13}{\atoms\per\centi\meter\squared} to \linebreak \SI{2E15}{\atoms\per\centi\meter\squared} were produced by MP. The investigations by alpha spectrometry, RI and SEM showed no significant differences of the sources in this range of areal densities. The area in contact with the electrolyte was completely covered with active material, but differences in homogeneity occurred (see Fig.~\ref{fig:3}~(b)). These were not correlated with the concentration, substrate or electrolyte. The SEM picture shows a clearly visible contrast of the deposited material but a surface structure similar to the one of the pristine substrate. This is in agreement with measurements of A. Vascon et al. \cite{Vascon2013}, who found the surfaces of thin sources on titanium-coated silicon wafers to be smoothest. The substrate roughness has a significant impact on the alpha spectra, as visible in Fig.~\ref{fig:12} as an increase of the tailing $\tau_1$ with increasing RMS roughness. Therefore, a smooth surface should be preferred for the source fabrication by MP. Furthermore, the alpha lines of the MP sources are well-resolved (see Fig.~\ref{fig:3}), with average $\sigma_1$ of \SI[separate-uncertainty = true]{7.7(16)}{\kilo\electronvolt}. The average fit parameter $\eta_1$ is very close to 1 (\SI[separate-uncertainty = true]{0.95(1)}{}) and the average tailing parameter $\tau_1$ is \SI[separate-uncertainty = true]{6.5(10)}{\kilo\electronvolt}, which is one of the lowest values when compared with those of the samples produced by the other fabrication methods (see Fig.~\ref{fig:9}). These results are in good agreement with AASI simulations of \textsuperscript{233}U sources with less than 20 atomic layers (see Fig.~\ref{fig:11}). In summary, the MP sources show very good properties; with probably less than 20 atomic layers they should reach a recoil efficiency of \textsuperscript{229}Th close to \SI{100}{\percent}. Nevertheless, the investigation of the recoil efficiency from a \textsuperscript{232}U source only showed an efficiency of \SI[separate-uncertainty = true]{10.5(11)}{\percent}. This discrepancy to theoretical suggestions based on the \textsuperscript{233}U experiments may have different reasons. First, the deposition of the \textsuperscript{232}U was visible with the naked eye in contrast to all of the \textsuperscript{233}U sources. Therefore, the deposited layer must be much thicker with lesser uranium atoms. This was checked by AASI simulations and comparison with a qualitative 4k alpha spectrum of the MP source. These AASI fits (with an areal uncertainty of \SI{0.7}{\percent} and a $\chi^2$ of 1.749) gave only a thickness of about \SI{35}{\nano\meter} with a RMS roughness of about \SI{90}{\nano\meter} and a density of \SI{6.25}{\gram\per\centi\meter\cubed} for the \textsuperscript{232}U MP source. The unexpected high density can only be caused by metallic, inactive impurities, since the total amount of all determined \textsuperscript{232}U and daughter atoms give a far lower weight (\SI{0.04}{\micro\gram} of a total simulated amount of \SI{10.9}{\micro\gram}, including also lighter atoms like hydrogen and oxygen). These impurities might be palladium from the anode or other metals like iron, tin or zinc, which were not investigated beforehand. Another reason for the discrepancy could be sputter effects. The high activity of the MP source corresponds to a high recoil ion rate. High energetic recoil ions, e.g. \textsuperscript{208}Pb from the alpha decay of the daughter nuclide \textsuperscript{212}Po ($E_{kin} = $\SI{166.9}{\kilo\electronvolt}), may also ablate collected \textsuperscript{228}Th from the catcher foil. This has to be investigated by further simulations before a final statement on the recoil efficiency of the MP sources can be made. The energy loss of the recoil ions is difficult to estimate, since the exact chemical composition of the MP sources is not known and, e.g., organic impurities from the electrolyte, which do not substantially affect alpha particles, or impurities by daughter nuclides could have a significant impact on the recoil ion energy.

\subsection{Drop-on-Demand inkjet printing}

The DoD printed sources were very thin with areal densities in the range of \SIrange[range-units=single, range-phrase= --]{1}{7E14}{\atoms\per\centi\meter\squared}. The deposits were not visually observable. SEM analysis was quite challenging, because the deposits were barely visible due to the height contrast of the titanium foils. The shape of the deposits on different regions of the titanium foils varied, probably due to differences in the surface properties. This also had an impact on the fit parameters $\eta_1$ and $\tau_1$. Some spectra of sources with large deposits had smaller values of the fit parameters (see DoD1, DoD2, DoD4 and DoD6 in Table~\ref{tab11}), whereas others with smaller deposits showed broader alpha peaks (see DoD3, DoD5 and DoD7 in Table~\ref{tab11}). The broadening is probably an effect of the more convex lateral shape of the deposits, as indicated by AASI simulations (Fig.~\ref{fig:13}). In dependence on the surface properties (hydrophilic or hydrophobic) of the backing, the lateral cut of deposits of evaporated drops have a concave or convex shape \cite{Haas2017}. The recoil efficiency for \textsuperscript{229}Th might be close to \SI{100}{\percent}, because the sources do not contain contamination of light elements in contrast to sources prepared by MP. Also the macroscopic areal distribution of the activity is homogeneous as shown in the RI in Fig.~\ref{fig:4}. The energy loss of the \textsuperscript{229}Th recoil ions can be estimated more easily than for MP sources, because the chemical species, uranyl nitrate, is known. As the number of atomic layers is estimated to be below 20 from simulated spectra (Fig.~\ref{fig:11}) compared with the experimental alpha spectrum in Fig.~\ref{fig:4}, the maximum energy loss of the \textsuperscript{229}Th recoil ions might be about \SI[separate-uncertainty = true]{9(2)}{\kilo\electronvolt} based on the values of Table~\ref{tab1}.

\subsection{Chelation by sulfonic acid groups}

The sources fabricated by chelation have the best quality by far as judged from the value of $\eta_1$ of about 1.0. The silicon surface provides a perfectly smooth surface for the sulfonic acid groups and therefore for the chelated \textsuperscript{233}U. This can be seen both in the SEM picture and in the RI by a very homogeneous areal distribution of the activity (see Fig.~\ref{fig:5}). The RI and also the areal \textsuperscript{233}U density indicate, though, that the densest occupation of the silicon surface by functional groups of about \SI{E15}{\hydroxygroup\per\centi\meter\squared} \cite{Aswal2006} has not yet been reached. The reason for the low areal density below \SI{E15}{\atoms\per\centi\meter\squared} could be due to two things. On one hand, preventing the formation of functional multilayers was problematic during the functionalization of the silicon surfaces. Therefore, conservative fabrication conditions were selected \cite{Aswal2006}. These may have led to fewer sites than needed to reach the theoretically possible limit of \SI{E15}{\atoms\per\centi\meter\squared}. The optimal conditions for these silicon surfaces would have to be approached iteratively. On the other hand, the second step of the functionalization, the oxidation of the thiol groups to sulfonic acid groups, is also complex for extremely thin layers. The process is quite delicate, as a specific concentration of hydrogen peroxide is required to oxidize the thiols to sulfonic acid groups. If the concentration is too low or the contact time too short, the thiol groups on the surfaces will be only partially oxidized and there might an intermediate stage containing disulfide bridges as a possible reaction product. If the hydrogen peroxide concentration is too high, the intermediate disulfide bridges are oxidized to sulfonic acid groups, but the functional layers are also destroyed. Still, the alpha spectra, as judged by the corresponding fit parameters, are the best when compared with those from sources fabricated by any other method. The fit routine had to be performed with a fixed $\eta_1 = 1$, because the multiplet peaks are too narrow and do not have a long tailing (parameterized by $\tau_2$) at all. A problem with the alpha spectra was the very low statistics due to the low areal uranium content, which is reflected in the larger error bars of $\sigma_1$ and $\tau_1$. The \textsuperscript{233}U multiplet in the alpha spectra is even better resolved than suggested by the AASI simulation (see Fig.~\ref{fig:11}, black line). All this indicates that the sources fabricated by chelation might have an almost \SI{100}{\percent} recoil efficiency of \textsuperscript{229}Th recoil ions. We estimate the energy loss to just a few \SI{}{\electronvolt}, due to breaking of chemical bonds.

\subsection{Self-adsorption}

The parameter studies for source fabrication by SA show optimal conditions at a pH value of 5 and on titanium foils thermally oxidized at \SI{500}{\degreeCelsius} for \SI{1}{\hour}. In this way, a maximum areal density of about \SI{80}{\nano\gram\per\centi\meter\squared} (\SI{2E14}{\atoms\per\centi\meter\squared}) was reached. The findings of uranium adsorption on anatase fit well with results reported in \cite{Lamb2016}. The evolution of the adsorption rate as a function of treatment temperature of the titanium foils (Fig.~\ref{fig:6}) cannot be explained by an increase of the RMS roughness, as the AFM data in Table~\ref{tab8} show. Moreover, the evolution shows the adjustment of the titanium modification on the foil surface in the variation of the uranium adsorption. Two peaks of adsorbed uranium are visible and correspond to anatase (at \SI{450}{\degreeCelsius}) and to rutile (at \SI{700}{\degreeCelsius}). At room temperature up to \SI{400}{\degreeCelsius}, amorphous titanium is present. Between \SI{450}{\degreeCelsius} and \SI{700}{\degreeCelsius}, a (semi-amorphous) mixture of both anatase and rutile is present. This fits well with x-ray diffraction measurements of sol-gel coated titanium films \cite{Velten2001}. It was possible to increase the amount of adsorbed uranium from \SI{80}{\nano\gram\per\centi\meter\squared} to \SI{270}{\nano\gram\per\centi\meter\squared} on titanium foils that were sand-blasted and then pretreated at \SI{450}{\degreeCelsius}. However, the roughness produced by sand-blasting was not well reproducible, as inferred from the large standard deviation of the amount of adsorbed uranium. The properties of sources prepared on untreated vs. sand-blasted Ti foils were very different. The areal distribution, shown on untreated foils in Fig.~\ref{fig:7}~(b), is very homogeneous. Only few artifacts are visible, probably caused by the quite rough surface of the titanium foils (see SEM picture in Fig.~\ref{fig:7}~(c)), as it was also observed on the titanium foils of the DoD sources (see Fig.~\ref{fig:4}). The three single peaks of the \textsuperscript{233}U multiplet are visible in the alpha spectra as shoulders of the next highest peak, due to the large average tailing $\tau_1$ of \SI[separate-uncertainty = true]{10.96(226)}{\kilo\electronvolt}, due to the rough surface of the titanium foil. The average $\tau_1$ value is similar to that of the DoD sources, but $\tau_2$ of \SI[separate-uncertainty = true]{50.35(539)}{\kilo\electronvolt} has a higher weighting factor due to a lower value of $\eta_1$ of \SI[separate-uncertainty = true]{0.75(6)}{\kilo\electronvolt} in the case of the untreated foils (see Fig.~\ref{fig:9}). The comparison with an AASI simulation in Fig.~\ref{fig:14}~(b) indicates that the reason for the differences in the peak shape is likely connected to the rough surface of the titanium foils. The RI and SEM pictures of the SA sources on sand-blasted foils show a much rougher surface and, therefore, more artifacts in the RI. Also the tailing in the alpha spectra is increased, leading to even lower values of $\eta_1$ of \SI[separate-uncertainty = true]{0.48(14)}{\kilo\electronvolt} and higher values of $\tau_1$ of \SI[separate-uncertainty = true]{14.76(1068)}{\kilo\electronvolt}. This supports the assumption that the tailing of the SA sources on untreated foils can be explained solely by the roughness of the foils. If the roughness of the substrates could be reduced, the SA sources would have the same quality as the sources fabricated by chelation with sulfonic acid groups concerning the recoil efficiency and the energy loss of \textsuperscript{229}Th recoil ions. However, the SA method provides currently a higher areal density. The investigation of the recoil efficiency from a \textsuperscript{232}U source produced on a limited area by SA gave an efficiency of \SI[separate-uncertainty = true]{94.2(33)}{\percent}, which fits very well with the theoretical suggestions. The small discrepancy to an optimum value of \SI{100}{\percent} might be caused by sputter effects due to high energetic recoil ions like \textsuperscript{208}Pb coming from the source. The spectra depicted in Fig.~\ref{fig:10}~(a) support the theoretical suggestion that the layers produced by SA are single atomic layers, since no \textsuperscript{232}U is visible in the spectrum of the catcher foil. In thicker deposits with several atomic layers, clusters of the source material are expected to be sputtered due to alpha decay collision cascades in deeper layers \cite{Nekrasov2020}. Furthermore, the areal distribution of \textsuperscript{232}U in the RI is as homogeneous as the distribution of \textsuperscript{233}U in the RI of Fig.~\ref{fig:7}~(b). Therefore, self-adsorption can also be carried out well on confined surfaces.

\section{Conclusion}\label{Concl}

Four different fabrication methods were used to produce \textsuperscript{233}U recoil ion sources delivering \textsuperscript{229(m)}Th in a most narrow kinetic energy distribution. They were analyzed both quantitatively and qualitatively with respect to the resolution of the alpha spectra and the areal density of the \textsuperscript{233}U layer. Simulations performed with AASI helped to understand the effects of different source characteristics like layer thickness, roughness, homogeneity and chemical species on the \textsuperscript{233}U alpha spectrum. Both the information from the simulations and the experimental alpha spectra were used to make predictions on recoil efficiency and energy loss of \textsuperscript{229}Th recoil ions. Additionally, experiments were performed with two \textsuperscript{232}U sources produced by MP and SA to investigate their \textsuperscript{228}Th recoil efficiency. Sources produced by chelation with sulfonic acid groups are likely the best recoil ion sources with respect to energy sharpness of the \textsuperscript{229}Th recoil ions. However, their areal density and hence the recoil ion rate is quite low. MP sources are very close in their quality to sources produced by chelation. However, they consist of several atomic layers and are therefore probably not ideal concerning the energy loss of recoil ions. This was proven with a \textsuperscript{232}U source, for which a recoil efficiency of roughly \SI{10}{\percent} was determined. Effects by metallic, inactive contaminants in the source material, which have caused a greater density in the deposit layer, have to be investigated by, e.g., neutron activation analysis in further experiments. They have to be removed before source production to achieve a higher recoil efficiency with MP sources. Also the methods of DoD and SA yielded promising sources. In contrast to MP and sources produced by chelation, they were fabricated on quite rough substrates. This induces a more pronounced tailing in the alpha spectra and decreases their quality. A method to control the shape of deposits produced by DoD is, e.g., to print on hydrophilic or even superhydrophilic surfaces. Such surfaces can be fabricated by anodic oxidation of titanium foils or by sol-gel coating of TiO$_2$ particles \cite{Velten2001}. The deposits would then result in very broad and regular shapes and therefore the quality of DoD sources could probably be improved. The suggested method will also improve the quality of the SA sources. After the implementation of these improvements, the sources produced by chelation and by self-adsorption may well be the best recoil ion sources due to their guaranteed monolayer thicknesses. The investigation of a \textsuperscript{232}U source produced by SA showed actually a \textsuperscript{228}Th recoil efficiency close to \SI{100}{\percent}, which fits perfectly to the proposed theoretical efficiency by alpha spectrometry analysis. MP and DoD would produce high qualitative recoil ion sources, but with a source thickness that can be less well controlled, and may, therefore, result in a larger energy spread and lower recoil efficiency of the \textsuperscript{229}Th ions. All the presented data fit well to the experiments and theoretical suggestions of Pohjalainen et al. \cite{Pohjalainen2020}, where sources with much larger areal uranium densities of \SIrange{74E15}{530E15}{\atoms\per\centi\meter\squared} were investigated and recoil efficiencies of about \SI{16}{\percent} were measured. For the TACTICa experiment, the fabricated sources by SA appear well suited. The actual energy distribution of the \textsuperscript{229}Th ions can be measured in the TACTICa ion source setup \cite{Haas2020} to gain further insights into the characteristics of the produced sources.

\section{Acknowledgement}
	
	The authors acknowledge the local support of the mechanical workshop and the electronics workshop at the TRIGA site of the Department of Chemistry at JGU Mainz. We also acknowledge the help of P. Th\"orle-Pospiech with the \textsuperscript{232}U separation. This work is supported by the Helmholtz Excellence Network ExNet020, Precision Physics, Fundamental Interactions and Structure of Matter (PRISMA$^+$) from the Helmholtz Initiative and Networking Fund.


\section*{Appendix}\label{App}
\begin{table*}[!htb]
	\caption{Fit parameters of the qualitative alpha spectra of all fabricated sources, sorted by the fabrication method, and of some simulated spectra. Constants are marked with a $*$ inside the error brackets because their error could not be determined. When $\eta_1$ was set to 1.0, the peak had no significant tailing and therefore had to be fitted only with $\tau_1$ and $\sigma_1$. As the second term of the fit function is multiplied with $\eta_2$=0, $\sigma_2$ and $\tau_2$ can be neglected. ``$\text{DoF}$'' is short for ``degrees of freedom''.}
	\label{tab11}
	\centering
	\begin{tabulary}{1.0\linewidth}{ L | C | C | C | C | C | C | C | C }
		\hline
		\multicolumn{9}{c}{Experimental spectra} \\
		\hline
		\mbox{sample} & \mbox{channels} & $\eta_1$ & $\sigma_1$ / keV & $\sigma_2$ / keV & $\tau_1$ / keV& $\tau_2$ / keV & $\chi^2/\text{DoF}$ & R$^2$ \\ 
		\hline 
		MP1 & 2048 & 1.00(*) & 7.51(137) & -- & 15.4(51) & -- & 1.55 & 0.7131\\ 
		MP2 & 2048 & 0.94(1) & 5.28(9) & 12.5(21) & 5.21(21) & 49.7(50) & 0.74 & 0.9965\\ 
		MP3 & 2048 & 1.00(*) & 10.6(21) & -- & 2.4(107) & -- & 1.41 & 0.8370\\ 
		MP4 & 2048 & 0.96(1) & 9.63(31) & 104(52) & 6.32(91) & 54.4(235) & 0.75 & 0.9884\\ 
		MP5 & 2048 & 0.96(1) & 8.13(22) & 84(32) & 8.04(56) & 65.1(173) & 0.68 & 0.9915\\ 
		\hline
		DoD1 & 2048 & 0.96(1) & 7.13(23) & 93.2(590) & 11.6(7) & 95.4(381) & 0.72 & 0.9897\\ 
		DoD2 & 1024 & 0.95(1) & 6.57(19) & 30.0(*) & 9.94(45) & 125(14) & 5.75$\text{E-6}$ & 0.9911\\ 
		DoD3 & 1024 & 0.72(6) & 5.92(38) & 12.3(15) & 15.8(16) & 63.0(57) & 3.52$\text{E-6}$ & 0.9881\\
		DoD4 & 2048 & 0.93(2) & 9.27(55) & 93.2(*) & 8.29(152) & 72.3(438) & 1.09 & 0.9660\\
		DoD5 & 2048 & 0.93(2) & 6.31(48) & 93.2(*) & 19.9(19) & 80.3(475) & 0.996 & 0.9662\\
		DoD6 & 1024 & 0.96(1) & 7.04(24) & 75.0(*) & 7.98(52) & 259(84) & 2.73$\text{E-6}$ & 0.9872\\
		DoD7 & 2048 & 0.67(7) & 8.34(40) & 14.2(12) & 11.0(20) & 52.9(52) & 0.97 & 0.9903\\
		\hline
		Ch1 & 2048 & 1.00(*) & 8.55(83) & -- & 7.95(206) & -- & 3.58 & 0.8840\\ 
		Ch2 & 1024 & 1.00(*) & 10.9(14) & -- & 5.12(334) & -- & 2.72 & 0.7269\\ 
		Ch3 & 1024 & 1.00(*) & 8.73(140) & -- & 9.32(344) & -- & 3.40$\text{E-6}$ & 0.8499\\
		Ch4 & 1024 & 1.00(*) & 10.1(22) & -- & 3.75(682) & -- & 5.87$\text{E-6}$ & 0.7034\\
		Ch5 & 1024 & 1.00(*) & 6.86(106) & -- & 11.1(22) & -- & 5.15$\text{E-6}$ & 0.8578\\
		Ch6 & 2048 & 1.00(*) & 9.24(96) & -- & 9.38(247) & -- & 1.70 & 0.8974\\
		\hline
		SA1 & 2048 & 0.75(6) & 10.8(4) & 5.00(258) & 11.1(22) & 50.7(56) & 0.92 & 0.9912\\
		SA2 & 2048 & 0.75(6) & 9.79(48) & 16.8(17) & 10.8(24) & 50.0(52) & 1.38 & 0.9886\\
		SA3 & 2048 & 0.62(3) & 12.6(10) & 30.1(43) & 25.4(56) & 215(57) & 0.90 & 0.9800\\
		SA4 & 2048 & 0.33(9) & 8.17(601) & 8.92(569) & 4.08(300) & 75.7(45) & 3.80 & 0.9798\\			
		\hline\noalign{\smallskip} 
		\hline
		\multicolumn{9}{c}{Simulated spectra} \\
		\hline
		L1 & 2048 & 0.91(1) & 4.8(1) & 6.6(21) & 10.1(3) & 65.7(44) & 1.18 & 0.9973\\
		L30 & 2048 & 0.96(3) & 9.1(3) & 11.0(*) & 18.7(19) & 116(74) & 6.93 & 0.9846\\
		L50 & 2048 & 0.94(*) & 14.4(10) & 17.5(*) & 18.5(121) & 96.5(491) & 33.2 & 0.9297\\
		\hline
		R0 & 2048 & 0.91(1) & 4.8(1) & 6.6(21) & 10.1(3) & 65.7(44) & 1.18 & 0.9973\\
		R17 & 2048 & 0.94(*) & 6.7(1) & 6.8(*) & 16.1(6) & 95.0(67) & 2.05 & 0.9954\\
		R51 & 2048 & 0.95(2) & 10.2(3) & 13.5(*) & 24.9(16) & 124(62) & 4.47 & 0.9896\\
		\hline
		norm & 2048 & 0.91(1) & 4.8(1) & 6.6(21) & 10.1(3) & 65.7(44) & 1.18 & 0.9973\\
		conv & 2048 & 0.93(*) & 7.0(1) & 7.0(*) & 12.5(4) & 86.5(43) & 1.98 & 0.9970\\
		conc & 2048 & 0.91(*) & 5.9(2) & 5.9(*) & 12.5(5) & 67.6(38) & 1.65 & 0.9933\\
		\hline
	\end{tabulary}
\end{table*}
\end{document}